\documentclass[12pt]{article}
\usepackage{xcolor,graphicx}
\usepackage[title]{appendix}
\usepackage{units,nicefrac,bm,amsmath,amssymb}
\usepackage{psfrag}
\usepackage[numbers,sort&compress]{natbib}
\usepackage{correct}
\begin{document}

\title{A multiscale modeling study for the convective mass transfer in a subsurface aquifer}

\author{Jahrul M Alam\\
  Department of Mathematics and Statistics, Memorial University \\
}

\date{Received: date / Accepted: date}

\maketitle

\begin{abstract}
Quantitative and realistic computer simulations of mass transfer associated with CO$_2$ disposal in subsurface aquifers is a challenging endeavor. This article has proposed a novel and efficient multiscale modeling framework, and has examined its potential to study the penetrative mass transfer in a CO$_2$ plume that migrates in an aquifer. Numerical simulations indicate that the migration of the injected CO$_2$ enhances the vorticity generation, and the dissolution of CO$_2$ has a strong effect on the natural convection mass transfer. The vorticity decays with the increase of the porosity. The time scale of the vertical migration of a CO$_2$ plume is strongly dependent on the rate of CO$_2$ dissolution.  Comparisons confirm the near optimal performance of the proposed multiscale model. These primary results with an idealized computational model of the CO$_2$ migration in an aquifer brings the potential of the proposed multiscale model to the field of heat and mass transfer in the geoscience.

{\bf keywords:}{mass transfer \and multiscale modeling \and porous media \and CO$_2$ plume \and subsurface flow \and aquifer \and wavelet method \and numerical modeling.}
\end{abstract}

\begin{tabular}{ll}
  \multicolumn{2}{l}{\bf List of symbols}\\
  $C$ & Concentration of CO$_2$\\
  $C_0$ & Reference concentration of CO$_2$\\
  $D$ & Diffusion coefficient ($\unit{m^2s^{-1}}$)\\
  $d$ & Microscopic length scale ($\unit{m}$)\\
  $g$& Acceleration due to gravity ($\unit{ms^{-2}}$)\\
  $H$ & Vertical length scale of the aquifer ($\unit{m}$)\\
  $K$ & Permeability ($\unit{m^2}$) \\
  $L$ & Horizontal length scale of the aquifer ($\unit{m}$) \\
  $M$ & Molar mass of CO$_2$ ($\unit{kg/mol}$)\\
  $N$ & Number of nodes on a regular mesh\\
  $\mathcal N$ & Number of nodes on an adaptive mesh\\
  $N_b$ & Buoyancy frequency ($\unit{s^{-1}}$)\\
  $\mathcal Da$ & Darcy number, $\frac{K}{H^2}$\\
  $\mathcal Re$ & Reynolds number at macroscale, $\frac{UH}{\nu}$\\
  $\mathcal Re_d$ & Reynolds number at pore scale, $\frac{Ud}{\nu}$\\
  $\mathcal Sc$ & Schmidt number, $\frac{\nu}{D}$\\
  $u_{D_i}$ & Space-time mean Darcian velocity ($\unit{ms^{-1}}$)\\
  $\langle\overline u_i\rangle$& Space-time mean intrinsic velocity ($\unit{ms^{-1}}$)\\
  $u'_i$ & Deviation from $u_{D_i}$ in a REV ($\unit{ms^{-1}}$)\\
  $u''_i$ & Deviation from $\langle\overline u_i\rangle$ in a REV ($\unit{ms^{-1}}$)\\
  $u_1,u_3$ & Dimensionless space-time mean intrinsic velocity\\
  $c$ & Dimensionless space-time mean intrinsic concentration\\
  $x_i$ & Cartesian coordinate in the $i$-th direction\\
  & \\
  \multicolumn{2}{l}{\bf Greek symbols}\\
  $\alpha$& Ratio of molecular viscosity to effective viscosity\\
  $\beta$& Solutal expansion coefficient\\
  $\Gamma$ & Rate of CO$_2$ dissolution ($\unit{Mm^{-1}}$)\\
  $\Delta V$ & Representative elementary volume~(REV)\\
  $\Delta V_f$ & The fraction of REV occupied with fluid\\
  $\Delta t$ & time step \\
  $\Delta x_i$ & Local step size in the $x_i$ direction\\
  $\Delta x$ & Minimum of $\Delta x_i$ for $i=1,\,3$\\
  $\phi \Del{= \frac{\Delta V_f}{\Delta V}}$ & Porosity \Add{$\left(\frac{\Delta V_f}{\Delta V}\right)$}\\
  $\rho_0$& Reference density \\
  $\lambda$ & Characteristic length scale of a REV \\
  $\epsilon$ & Error tolerance \\
  $\mu$ & Molecular viscosity ($\unit{kgs^{-1}m^{-1}}$)\\
  $\nu$ & Kinematic viscosity ($\unit{m^2s^{-1}}$)\\
\end{tabular}

\section{Introduction}
\begin{figure}[hb]
  \centering
  \begin{tabular}{cc}
    \includegraphics[trim=1.5cm 0cm 2.5cm 0cm,clip=true,width=4.5cm,height=3.5cm]{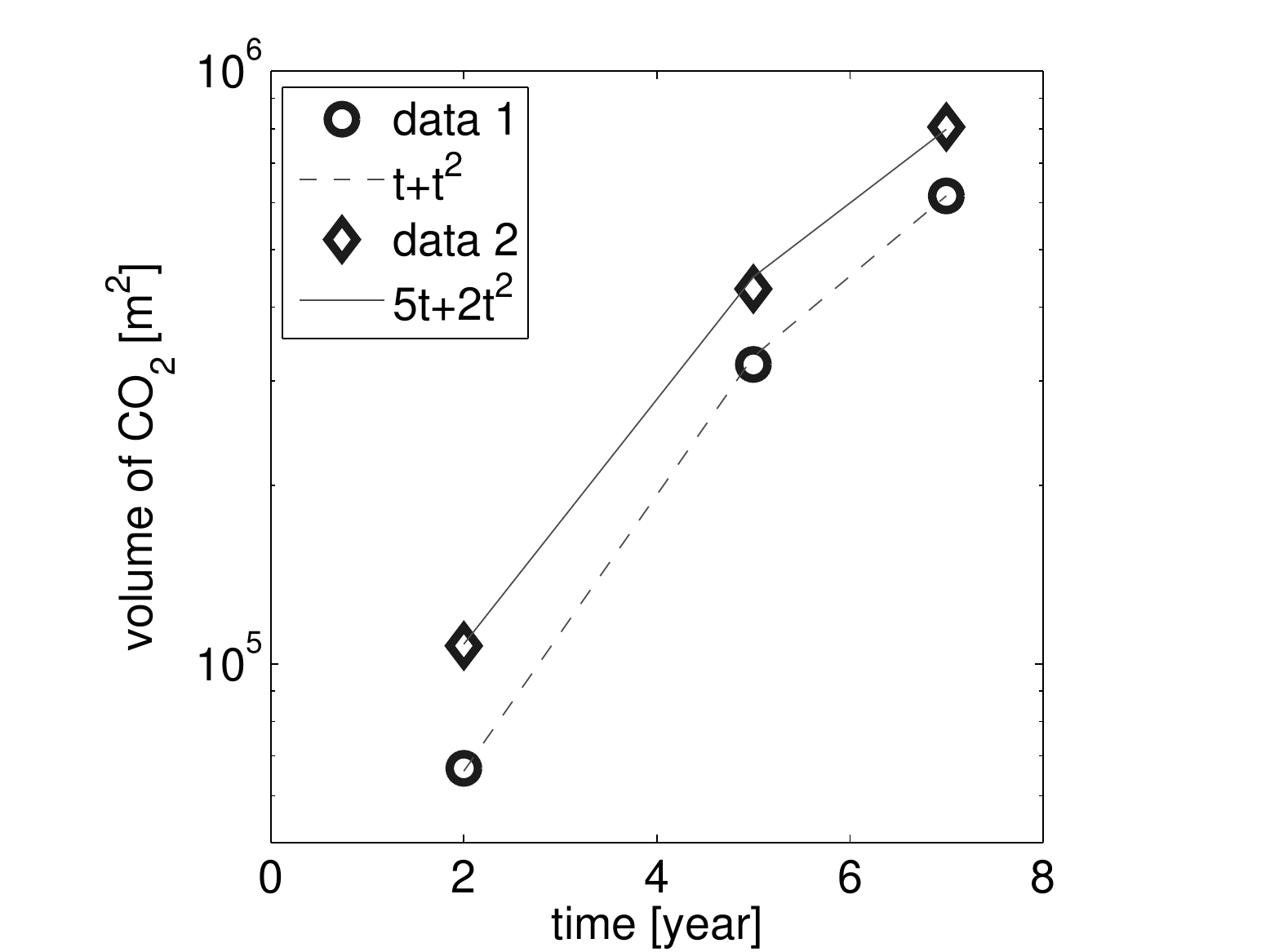}&
    \includegraphics[trim=-1.75cm 0cm 0cm 0cm,clip=true,width=3.25cm,height=3.5cm,angle=90]{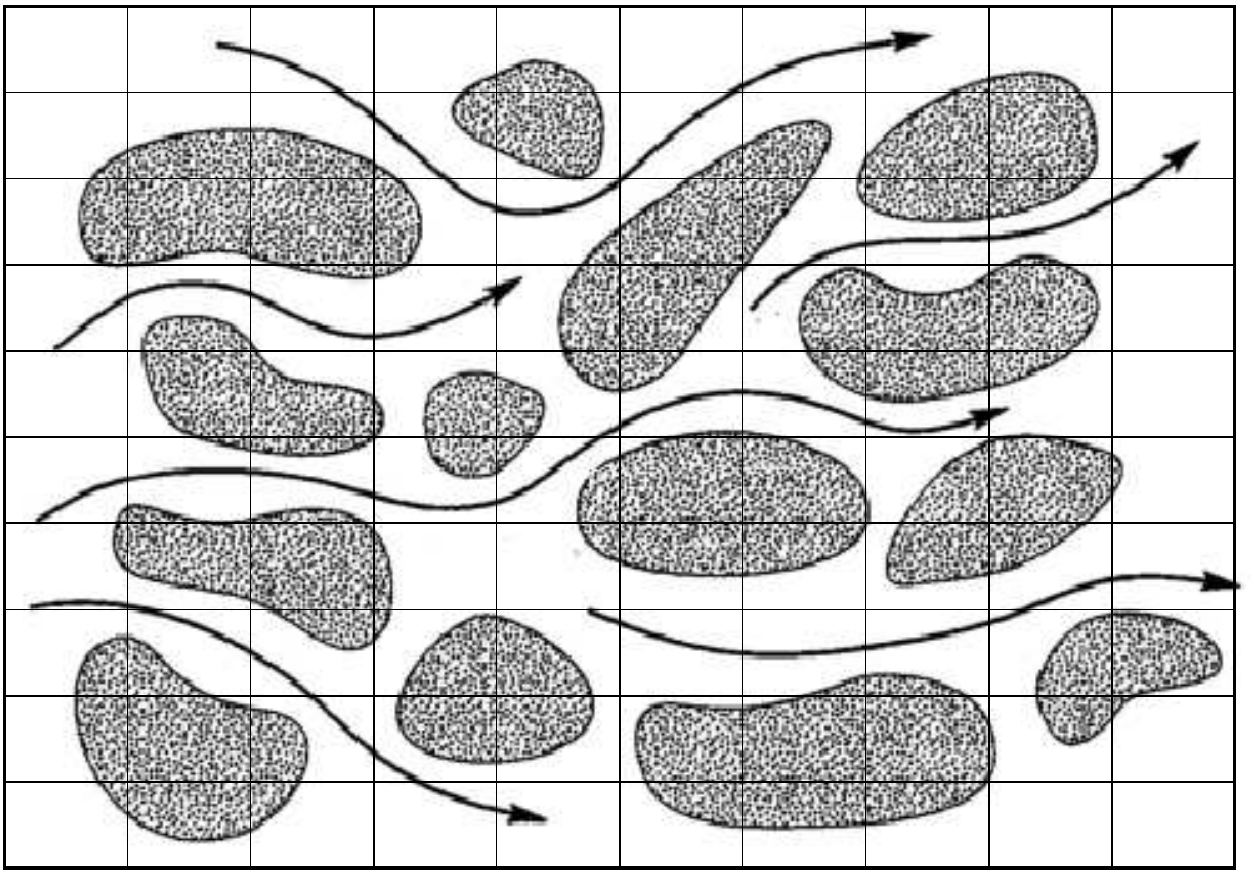}\\
    $(a)$ & $(b)$
  \end{tabular}
  \caption{$(a)$~Observed volumes of CO$_2$ near the top of the Sleipner aquifer between $1999$ and $2006$; {\tt data 1}:~volume ($\unit{m}^3$) of CO$_2$ by method $1$, broken line: fitted curve, $t+t^2$, {\tt data 2}:~volume ($\unit{m}^3$) of CO$_2$ by method $2$, and solid line: fitted curve, $5t+2t^2$~({\em e.g.}~\cite{Chadwick2009}). A parabolic increase of CO$_2$ volume as a function of time, $t$~[year] is noticed. $(b)$~A schematic representation of pathlines for plumes in an aquifer. Clearly, grid points may be placed on rocks or very low permeability zone, where a high resolution would place grid points close to pathlines, thereby improving the accuracy.}
  \label{fig:co2flux}
\end{figure}

The natural and mixed convection heat and mass transfer is an important topic of scientific scrutiny in a subsurface flow problem that investigates the disposal of anthropogenic CO$_2$ from the atmosphere into saline aquifers~\cite{Nordbotten2005}.
Example of such projects include the Utsira sand at Sleipner, Norway~\cite{Chadwick2009}, the Mt. Simon aquifer in the Illinois basin~\cite{Leetaru2009,Bachu94}, saline aquifers in the Alberta basin, Canada~\cite{Pruess2008}, and the Carrizo-Willcox aquifer in Texas~(CWT)~\cite{Nicot2008}. Seismic data from the Sleipner shows a marked increase in the CO$_2$ flux near the reservoir top~(see Fig~\ref{fig:co2flux}$a$ of~\cite{Chadwick2009}), which suggests further investigation of the rate of vertical migration and progressive development of CO$_2$ plumes~({\em e.g.}~\cite{Chadwick2009}). \citet{Chadwick2009} hypothesized that this marked variability may be due to the multiscale natural convection mass transfer associated with the non-Darcian plume migration through numerous pathways in the aquifer. In the same vein, multiscale processes associated to a plume migration in the Carrizo-Willcox aquifer was investigated numerically by~\citet{Pruess2011} using a classical macroscopic Darcian model along with a sub-grid scale parameterization scheme (albeit the exact form of the scheme was not outlined with full details in~\cite{Pruess2011}). Clearly, a complete understanding of the multiscale mass transfer mechanism in aquifers remains an active research area in the field of geoscience and reservoir engineering~\cite{Nicot2008,Pruess2011,Lindeberg97,Riaz2006,Farajzadeh2007,Pau2010,Johannsen2012}. Hence, there are increasing interests on extending the adaptive mesh and multiscale finite volume/element methods (AMR based methods) for high performance numerical simulations of flow and transport in saline aquifers ({\em e.g.,} see,~\cite{Farajzadeh2007,Hesse2008,Kunze2013,Nithiarasu99,Nordbotten2009,Jenny2006,Pau2010,Johannsen2012}). 

The present article has investigated the development of a novel adaptive wavelet multiscale modeling and simulation methodology for studying the non-Darcian flow and transport through aquifers. One objective of this article is to investigate an effective methodology to minimize computation work and to improve the accuracy for transport problems in aquifers that deal with multiscale phenomena, in comparison to the commonly used classical numerical methods. For example, some authors verified with a classical method that a $\Delta x$ between $5\times 10^{-4}~\unit{m}$ and $10^{-3}~\unit{m}$ is necessary to capture multiscale features ({\em e.g.}, see~\cite{Pruess2008,Pruess2011,Pau2010}), for which, the number of the grid points is about $N=10^{13}$ in the domain $1~\unit{m}\times 5~\unit{m}$~({\em e.g.}~\cite{Pruess2008}), and hence, $N$ would be far beyond the limit of modern computers for such a faithful simulation in a vertical cross section ($100~\unit{km}\times 200~\unit{m}$) of the CWT~\cite{Pruess2011}. In contrast, this article investigates a multiscale methodology to put the computational effort locally in the physical domain  (see~\cite{Oleg2005,Alam2012,Alam2014,Alam2014b,Wirasaet2006,Wirasaet2005b} for a technical details), where it is  necessary to adapt the computational work to field-scale geological structures and complex physical processes in aquifers. 

In the same vein, AMR based classical adaptive and mutiscale methods require to find an individual error monitoring scheme for each new applications ({\em e.g.}~\cite{Jenny2006,Kunze2013,Pau2010}). These methods were developed for steady elliptic problems, and their extension to multiscale flow and transport in aquifers is being investigated by a number of other authors (see the recent work of~\citet{Kunze2013}). 
Note also that an extremely small time step ($\Delta t$) is needed when these AMR based methods explicitly simulate natural and mixed convection heat and mass transfer in aquifers. In contrast, the present article investigates a multiscale methodology that captures multiscale physics with a more robust wavelet based computation, where the error is controlled {\em a priori} according to a prescribed tolerance~($\epsilon$)~\cite{Wirasaet2006,Wirasaet2005b}, and the time step ($\Delta t$) can be chosen independent of the spatial grid~($\Delta x$)~\cite{Alam2012,Alam2014}. In addition, the present research employs the volumetric mean of the time-averaged conservation laws with respect to a representative elementary volume~(REV) so that the non-Darcian multiscale features ({\em e.g.}~\cite{Chadwick2009}) in a flow may be resolved with a non-Darcian multiscale model  (see~\cite{Whitaker99,DeLemos2006} and the refs therein). This approach provides a natural framework for adopting appropriate parameterization for phenomena which are not directly resolved with the mean conservation laws~({\em e.g.}~\cite{Pruess2011}). Here, the effect of CO$_2$ dissolution on the gravitational segregation of a CO$_2$ plume has been parameterized. Most importantly, the size of the REV has been adapted dynamically to the local variation of the CO$_2$ mass fraction. To the best of knowledge, the present development is a first time investigation for the multiscale natural convection mass transfer in a saline aquifer, and it complements the growing trend in computational modeling of the natural/mixed convection phenomena in subsurface flows as well as in other fields~\cite{Jenny2006,Nordbotten2009,Pau2010,Popiolek2009,Wirasaet2006,Wirasaet2005b}. Following are few remarks on the proposed multiscale methodology.

\paragraph{Remarks:}

\begin{itemize}
\item The flow and transport at the reservoir scale is approximated with the classical Darcian approach. The multiscale mass transfer has been resolved with a non-Darcian approach. Subgrid scale parameterization schemes for both the momentum diffusion and the CO$_2$ dissolution have been adopted.   

\item A multilevel algorithm has been developed for the nonlinear coupling between the mass and momentum transfer processes in aquifers, which occurs at different temporal and spatial scales. 

\item Spatial differential operators are discretized using $\mathcal N$ significant wavelets, where each wavelet represents the change in scale near a grid point, and multiscale features are captured with a multiscale wavelet theory.

\item There are two important computational benefits. First, the number of grid points $\mathcal N$ is significantly small compared to the number of grid points needed for a classical numerical method if the same level of accuracy is desired. Second, if $\mathcal N$ increases, then the CPU time increases approximately linearly. Note that for a given tolerance $\epsilon$, $\mathcal N$ may increase due to change in gradients of solution. 

\item The error for such a simulation is controlled in both space and time according to the prescribed tolerance ($\epsilon$) and the maximum allowable CFL number, which is verified with a large number($>50$) of  numerical experiments.
\end{itemize}

Section~\ref{sec:multi} summarizes the new developments towards capturing the multiscale features in a subsurface plume migration. The adaptive multiscale methodology has been presented in section~\ref{sec:comp}. Representative results from a series of numerical experiments have been presented in section~\ref{sec:ver}. Finally, section~\ref{sec:sum} discusses the potential extension of the present development in the field of computational heat and mass transfer analysis.

\section{The proposed multiscale modeling framework}\label{sec:multi}
\Add{This model aims to simulate the natural convection mass transfer phenomena associated with CO$_2$ storage and migration in a typical aquifer, such as the Carrizo-Wilcox aquifer, Texas, US (a further details in section~\ref{sec:CWT}). In the following development, the aquifer height, $H$, is the length scale, and a typical flow speed, $U$, in the aquifer is the velocity scale. A vertical cross section of the aquifer is assumed to be simulated, which is a good approximation to represent overall multiscale features of the problem(see,~\cite{Pruess2008,Pruess2011}). The boundary conditions assume impermeable caproks on the bottom and top boundaries as well as open lateral boundaries. The aquifer is assumed a homogeneous and isotropic porous medium.} 

The present model diagnoses the flow and transport at the reservoir scale from the classical Darcian approach. The multiscale mass transfer has been resolved with a non-Darcian approach, where the REVs are resized locally in order to adapt the effective numerical resolution to the localized features of mass transfer. Subgrid scale parameterization schemes for both the momentum diffusion and the CO$_2$ dissolution have been adopted.   

\subsection{Classical upscaling approach for diagnosing macroscopic transfer}
The pore scale~(fine scale) heat, mass, and momentum transfer is replaced to the reservoir scale (macroscale)~({\em e.g.}~\cite{Pau2010,Bear72}), where
$$
\overline u_i = \frac{1}{T_0}\int_t^{t+T_0} u_i(x_i,t')dt'
\quad\hbox{and}\quad
u_{D_i} = \langle\overline u_i\rangle^v \equiv \frac{1}{\Delta V}\int_{\Delta V} \overline u_i dV
$$
are the time and volume averages of the velocity field, respectively. The Darcian velocity $u_{D_i}$
is a space-time mean with respect to the REV~($\Delta V$) and the  time scale, $T_0$, which  contains both the solid and the fluid~(see~\cite{Bear72} for details). Therefore, the natural convection mass transfer in an aquifer is typically studied with the following upscaling representation~({\em e.g.}, \cite{Pau2010,Farajzadeh2009,Brian2008,Homsy87,Riaz2003,Peaceman62})
\begin{equation}
  \label{eq:darcy}
   u_{D_i} = -\frac{K}{\mu}\left(\frac{\partial p}{\partial x_i} +\rho(C) g\delta_{i3}\right), \quad\frac{\partial u_{D_i}}{\partial x_i} =0,
\end{equation}
\begin{equation}
  \label{eq:dc}
  \frac{\partial C}{\partial t} +\frac{\partial}{\partial x_j}\left(\frac{1}{\phi}u_{D_j} C - D\frac{\partial C}{\partial x_j}\right)=0.
\end{equation}
In this model, the momentum transfer is in a steady state,  both the advection $u_{D_j}\frac{\partial u_{D_i}}{\partial x_j}$ and the deviatoric stress make a negligible contribution to momentum transfer, the drag between the porous medium and the fluid is assumed linearly proportional to the pressure gradient, and the Reynolds number, $Re_d = \frac{\rho Ud}{\mu}$ does not exceed a value about $1$~({\em e.g.}, see Chapter 5,~\cite{Bear72}). 

As discussed in the introduction, using the macroscopic upscaling~(\ref{eq:darcy}-\ref{eq:dc}), faithful simulations of the convective mass transfer in an aquifer often requires a high numerical resolution~({\em e.g.}~\cite{Pruess2008,Pruess2011,Pau2010}) because the characteristic length scale ($\lambda$) of the REV is typically at the order of $\Delta x$. 
As demonstrated schematically in Fig~\ref{fig:co2flux}$(b)$, $\Delta x$ needs to be sufficiently small in order for $\lambda\rightarrow d$ so that computations on grid points would represent the actual mass transfer along streamlines. In the following two sections, a multiscale representation of the fine scale mass and momentum transfer has been studied, using the most robust volumetric mean of the time-averaged conservation laws~\cite{Bear72,Rees2006,Braga2009,Yang2012}.

\subsection{Multiscale intrinsic mass transfer model}
\label{sec:mimtm}
The intrinsic space-time average is
$$
\langle\overline u_i\rangle = \frac{1}{\Delta V_f}\int_{\Delta V}\overline u_i (x_i,t) dV,
$$
where $\Delta V_f$ is the volume of the fluid contained in a REV. Note the difference between $\langle\overline u_i\rangle^v$ and $\langle\overline u_i\rangle$, 
where $u_{D_i} = \phi\langle\overline u_i\rangle$~\cite{Hsu90}.
The time and volume averages are assumed to commute; {\em i.e,.}
$
\langle\overline u_i\rangle = \overline{\langle u_i\rangle},
$
one may apply the temporal average into a volume averaged variable, and {\em vice-versa}.
Using the intrinsic average $\langle\overline\cdot\rangle$, the space-time transmissivity in a REV may be resolved, which is neglected in the classical macroscopic model~(\ref{eq:darcy}-\ref{eq:dc}).

To illustrate the multiscale framework, we begin with the decomposition
$$u_i = u_{D_i} + u'_i + u''_i$$
at three scales,
where the first component, $u_{D_i}=\langle\overline u_i\rangle^v$, captures the mean discharge per unit area, but does not represent a details of the flow. The second component, $u'_i$, adds the missing details into $u_{D_i}$ such that $\langle\overline{u'}_i\rangle^v=0$ and $\langle\overline{u'}_i\rangle\ne 0$. The third component, $u''_i$, represents a further details that satisfies $\langle\overline{u''}_i\rangle^v=0$ and $\langle\overline{u''}_i\rangle=0$. Clearly, the space-time mean intrinsic velocity $\langle\overline u_i\rangle = u_{D_i}+u_i'$ resolves an additional details $u_i'$ with respect to the classical mean Darcy velocity $u_{D_i}$. Similarly, the space-time intrinsic mean of the concentration of CO$_2$ can be obtained. In order to simplify the symbolic representation, $u_i$ is used for the dimensionless mean velocity $\langle\overline u_i\rangle$, and $c$ for the dimensionless mean concentration $\langle\overline C\rangle$ in the rest of this article, where all quantities are assumed uniform in the $x_2$ direction. \Add{Note that this two-dimensional assumption is an idealization for the radial symmetry of the dynamics of the CO$_2$ plume~(e.g.,~\cite{Nordbotten2009,Pruess2011}), and has been adopted to aid the investigation on the proposed multiscale model development.} 

The derivation of the present intrinsic multiscale model for the mass transfer is similar to what was detailed by~\citet{Rees2006} for the heat transfer, and by~\citet{Lage2002} for the momentum transfer~(see also~\cite{Bear72,Braga2009,Yang2012,Hsu90,Nithiarasu99}). However, in the heat transfer model, an independent diffusion equation was considered in~\cite{Rees2006} to account for thermal conduction through the solid phase. In the present mass transfer model, the dissolution of the invaded phase has been parameterized through the averaging process. \Add{At a depth below $800$~m (see, \cite{Benson2005}), the geothermal effects may be compensated by the geo-pressure gradient (see,~\cite{Lindeberg97} and chapter~2 of~\cite{DeLemos2006}), which is further illustrated by~\citet{Pruess2011}. The pressure gradient terms in eqs.~(\ref{eq:hrmm}-\ref{eq:vrmm}) accounts for the geothermal pressure gradient. Since studies indicate a weak dependence between the molar volume of dissolved CO$_2$ and density of the binary mixture, the Boussinesq approximation is reasonable~\cite{Barbero83,Lindeberg97,DeLemos2006,Pau2010}.} Following the derivation of~\citet{Lage2002}, the macroscopic upscaling~(\ref{eq:darcy}-\ref{eq:dc}) has been replaced with the multiscale upscaling~(\ref{eq:inc2d}-\ref{eq:mnc}):
\begin{equation}
  \label{eq:inc2d}
\frac{\partial u_1}{\partial x_1} + \frac{\partial u_3}{\partial x_3} = 0
\end{equation}
\begin{equation}
  \label{eq:hrmm}
\frac{\partial u_1}{\partial t} + u_1\frac{\partial (u_1/\phi)}{\partial x_1} + u_3\frac{\partial (u_1/\phi)}{\partial x_3} = - \frac{\partial (P\phi)}{\partial x_1} +\frac{\alpha}{\mathcal Re}\nabla^2 u_1 - \frac{\phi u_1}{\mathcal Da\mathcal Re}  
\end{equation}
\begin{equation}
  \label{eq:vrmm}
\frac{\partial u_3}{\partial t} + u_1\frac{\partial (u_3/\phi)}{\partial x_1} + u_3\frac{\partial (u_3/\phi)}{\partial x_3} = - \frac{\partial (P\phi)}{\partial x_3} +\frac{\alpha}{\mathcal Re}\nabla^2 u_3 - \frac{\phi u_3}{\mathcal Da\mathcal Re} + \frac{\mathcal Gr}{\mathcal Re^2}c
\end{equation}
\begin{equation}
  \label{eq:mnc}
\frac{\partial c}{\partial t} + u_1\frac{\partial c}{\partial x_1} + u_3\frac{\partial c}{\partial x_3} = - \frac{\mathcal Re^2}{\mathcal Gr\mathcal Fr^2}u_3 +\frac{1}{\mathcal Re\mathcal Sc}\nabla^2 c.
\end{equation}
In this model, the Schmidt number is $\mathcal Sc= \frac{\nu}{D}$, the macroscopic Reynolds number and the Darcy number are defined by $\mathcal Re = \frac{\rho UH}{\mu}$ and $\mathcal Da = K/H^2$, respectively, where $\mathcal Re_d\ll\mathcal Re$ and $\mathcal Da = \mathcal O(1/\mathcal Re)$. Clearly, if $\mathcal Re\rightarrow\infty$, we have $\mathcal Da\rightarrow 0$, which corresponds to a macroscopic model~(\ref{eq:darcy}-\ref{eq:dc}). 

\Add{As described in Appendix~\ref{app:ddm}, the space-time average of the nonlinear advection of momentum takes the form
$$
\left\langle\overline{u_j\frac{\partial u_i}{\partial x_j}}\right\rangle = 
\langle\overline u_j\rangle\frac{\partial\langle\overline u_i\rangle}{\partial x_j}+\frac{\partial}{\partial x_j}\left(\overline{\langle u''_i\rangle\langle u''_j\rangle} 
+ \langle\overline{ u''_i u''_j}\rangle
+  \langle\overline u''_i\overline u''_j\rangle \right),
$$
}
\Del{In the present model, the average of the nonlinear advection of the momentum has }\Add{which }resulted into the additional term 
$$
\frac{\partial}{\partial x_j}\left(\overline{\langle u''_i\rangle\langle u''_j\rangle} 
+ \langle\overline{ u''_i u''_j}\rangle
+  \langle\overline u''_i\overline u''_j\rangle \right),
$$
of which, the first two components have been neglected in the present laminar mass transfer model, and the last term can be parameterized with the Brinkman model~\cite{Brinkman49}, which -- in the dimensionless form -- has taken the form of the second term on the right hand sides~(rhs) of both~(\ref{eq:hrmm}) and~(\ref{eq:vrmm}). 
The third term on the rhs of~(\ref{eq:hrmm}, \ref{eq:vrmm}) has appeared due to the intrinsic average of the viscous and pressure stress, which models the density of the pressure drag and skin friction, %
$$ f_i = -\frac{\mu}{K}u_{D_i},$$
when the Reynolds number is small, {\em i.e.}, $\mathcal Re_d=\mathcal O(1)$~\citep{DeLemos2006}.

Similarly, the nonlinear advection of the mass flux has resulted into the additional term
$$
\frac{\partial}{\partial x_j}\left( \overline{\langle u''_i\rangle\langle c''_j\rangle} 
+ \langle\overline{ u''_i c''_j}\rangle
+  \langle\overline u''_i\overline c''_j\rangle \right)
$$
of which the first two components have been neglected and the last component has been parameterized to model the effect of dissolution with the first term on the right hand side of~(\ref{eq:mnc}).%

\subsection{The dissolution of CO$_2$ through the dispersive mass flux}
Literature review does not indicate a common approach for modeling the effect CO$_2$ dissolution into the resident saline, despite there are few attempts~({\em e.g.} see~\cite{Pruess2011}). If a CO$_2$ plume proceeds upward from an isolated source, the saline density at the interface increases by about $0.1$-$1\%$~\cite{Pau2010}. To a first order approximation, the global density of the background environment adopts a vertically decreasing profile when the plume migrates upward, albeit more specific observational data would confirm the actual density profile ({\em e.g.} see also the density distribution presented by~\citet{Pruess2011}).  

The present work has proposed a simple model
$$
\frac{\partial}{\partial x_j}
  \langle\overline u''_i\overline c''_j\rangle
= u_3\frac{\partial\tilde c}{\partial x_3}
$$
for the mass dispersion associated with the microscopic space-time variation of mass and momentum, where
a steady state horizontally homogeneous background concentration $\tilde c(x_3) = C_0+\Gamma x_3$ has been assumed
since the molecular mass of CO$_2$ is larger than that of brine~(see~\cite{Riaz2006,Hesse2008,Pau2010}).
With this model, one can define a Buoyancy frequency by $N^2_b = gM\Gamma/\rho_0$ to characterize the effect of dissolution, where $M$ is the molar mass of CO$_2$ and $\rho_0=C_0M$ is a reference density. Here, $N_b^2 >0$ corresponds to a situation with $\Gamma>0$; {\em i.e.}, in this case, CO$_2$ has been accumulated near the reservoir top, where the dissolution would result into gravitational fingers studied by~\citet{Pau2010}. $N_b^2<0$ corresponds to $\Gamma < 0$; {\em i.e.}, CO$_2$ dissolution is now associated with a vertically migrating plume~\cite{Chadwick2009}.
Accordingly, a Froude number is given by
$$\mathcal Fr^2 = \frac{U^2}{N^2_b H^2},$$ 
and using data from the Carrizo-Wilcox aquifer, Texas, {\em e.g.}, $U\sim 5\times 10^{-5}\unit{m/s}$ and $H\sim 200~\unit{m}$~\cite{Pruess2011,Nicot2008}, we estimate that $\mathcal Fr\ge 1$ corresponds to $N_b \le 2.5\times 10^{-7}~\unit{s^{-1}}$ and $\Gamma\ge -10^{-10}~\unit{M/m}$. Clearly, $\Gamma\ll -10^{-10}~\unit{M/m}$ results into $\mathcal Fr\ll 1$, and the mass transfer analysis for varying $\mathcal Fr$ exploits the effect of the CO$_2$ dissolution.

\subsection{The natural convection mass transfer as a function of $\mathcal Fr$}
The dominant mechanism for the onset of background dissolution on the natural convection mass transfer during the migration of a CO$_2$ plume has now been studied with a dimensional analysis, which explains the effect of the variation of the Froude number. %
To estimate the order of magnitude of each term in the mass and momentum conservation laws~\Add{(see~\citet{Rees2006})}, %
consider the solutal Grashof number,
$$Gr = \frac{g\Delta\rho H^3}{\rho_0\nu^2} = \frac{g\beta\Delta c H^3}{\nu^2},$$
the Schmidt number, $Sc=\nu/D$, the horizontal length scale, $L$, and the vertical length scale, $H$. 
\Add{Note the large aspect ratio of typical aquifers ({\em e.g.} the Carrizo-Wilcox aquifer, Texas, US~\cite{Nicot2008}) and the order of magnitude for the inertial term, $u_1\frac{\partial u_1}{\partial x_1} + u_3\frac{\partial u_1}{\partial x_3} = \mathcal O(\sqrt{\mathcal Gr}H^2/L^2)$.} 
Using horizontal and vertical length scales, 
$L\sim 10~\unit{km}$ and $H\sim 200~\unit{m}$, respectively~({\em e.g.}~\cite{Hesse2008}), where $H^2/L^2\ll 1$, as well as a fixed $Gr < 2\,500$, and in the limit of $\mathcal Da\rightarrow\infty$,  we obtain the following dimensionless linear system of PDEs
$$
\frac{\partial u_1}{\partial x_1} + \frac{\partial u_3}{\partial x_3}  = 0,
$$
$$0=-\frac{\partial p}{\partial x_1} + \frac{\partial^2 u_1}{\partial x_3^2},$$
$$0=-\frac{\partial p}{\partial x_3} + c,$$
$$u_3=\frac{\partial^2 c}{\partial x_3^2}.$$
The system is independent of the length and time scales, as well as of other dimensional parameters, and hence, exhibits a self-similar solution. The existence of a self similar solution indicates that  the vertical length scale can be determined from the dimensional parameters $\nu$, $D$, $H$, $L$, and $N_b$ those appear in the dimensional system of equations. A dimensional reasoning can be used to define the vertical length scale as a function of the remaining other parameters; {\em i.e.},
$$ H^6 = \frac{\nu D L^2}{N_b^2},$$
which gives an aspect ratio between the vertical scale and the horizontal scale:
$$\left(\frac{H}{L}\right)^4 = \frac{Fr^2}{GrSc}.$$ Clearly, in the limit of $\mathcal Fr\rightarrow\infty$ for a fixed $Gr$ and $Sc$, the vertical length scale extends to infinity; {\em i.e.}, $H\rightarrow\infty$. This indicates that a non-zero gradient, $\Gamma$, tends to reduce vertical migration of the CO$_2$ plume. For too small a value of $\mathcal Fr$, the vertical length scale is also too small; the plume will not continue to migrate vertically upward for small $\mathcal Fr$. This dimensional analysis has a good agreement with the numerical simulations presented in section~\ref{sec:Fr}.

This section concludes by noting that the analysis of the multiscale model~(\ref{eq:inc2d}-\ref{eq:mnc}) with an adaptive wavelet multiscale technique is a novel contribution of the present research, in contrast to classical models, where the macroscopic model~(\ref{eq:darcy}-\ref{eq:dc}) analysed with a multiscale technique~\cite{Pau2010,Jenny2006,Nordbotten2009}.  

\section{The adaptive wavelet multiscale simulation methodology}\label{sec:comp}

To outline the proposed numerical method, eqs~(\ref{eq:hrmm}-\ref{eq:mnc}) has been written in the following compact form:
\begin{equation}
  \label{eq:model}
  \frac{\partial\Psi_i}{\partial t} + u_j\frac{\partial\Psi_i}{\partial x_j} = \frac{\partial R_{ij}}{\partial x_j}
\end{equation}
where $\Psi_i$ represents $u_1$, $u_3$, or $c$, and $R_{ij}$ ($i,j=1,\,3$) represent the right hand sides of eqs~(\ref{eq:hrmm}-\ref{eq:mnc}). Since the velocity~($u_1,u_3$) and the concentration~($c$) depends on each other simultaneously in a natural convection mass transfer application, a fully implicit time integration scheme has been adopted, where $u_1$, $u_3$, and $c$ are computed simultaneously. %

\subsection{Time integration}
A fractional-step method, originally proposed by~\citet{Chorin68}, has been applied to solve~(\ref{eq:model}) for $\Psi_i^{n+1/2}$, %
$$
  \frac{\Psi_i^{n+\frac{1}{2}}-\Psi_i^n}{\Delta t} +\frac{1}{2}\left( u_j^{n+\frac{1}{2}}\frac{\partial\Psi_i^{n+\frac{1}{2}}}{\partial x_j} + u_j^{n}\frac{\partial\Psi_i^{n}}{\partial x_j} \right) = \frac{1}{2}\left( R_{ij}^{n+\frac{1}{2}} + R_{ij}^n\right),
$$
and it takes the following symbolic form
\begin{equation}
  \label{eq:bvp}
  \mathcal L(\Psi_i^{n+\frac{1}{2}}) = \bm S(\Psi_i^n),
\end{equation}
where superscripts $n$ and $n+\frac{1}{2}$ mean the present time step and the first fraction of the next time step, respectively.

In the second fraction of a time step, eq.~(\ref{eq:inc2d}) is satisfied, $\frac{\partial u_j^{n+1}}{\partial x_j} = 0$, such that the macroscopic model~(\ref{eq:darcy}) is approximately diagnosed from
\begin{equation}
  \label{eq:dp}
  -\frac{\partial}{\partial x_j}\left(\frac{\partial (P\phi)^{n+1}}{\partial x_j}\right)
  = \frac{1}{\Delta t}
  \frac{\partial}{\partial x_i}\left(u_{D_i}^{n+1}\right),
\end{equation}
where $u_{D_i}^{n+1} = u_i^{n+1}-u_i^{n+1/2}$. 
Setting $c^{n+1}=c^{n+\frac{1}{2}}$ from the nonlinear algebraic system~(\ref{eq:bvp}), $u_i^{n+1}$ is also diagnosed from the elliptic problem~(\ref{eq:dp}), where (\ref{eq:bvp}) and (\ref{eq:dp})   must be solved with efficient iterative solvers at each time step. 

Since~(\ref{eq:bvp}) is a nonlinear system, the classical Newton's method 
$$\Psi_i^{k+1,n+\frac{1}{2}} = \Psi_i^{k,n+\frac{1}{2}}+s_k$$
such that
$$\mathcal Js_k = \mathcal S(\Psi_i^n)-\mathcal L\left(\Psi_i^{k,n+\frac{1}{2}}\right)$$
must evaluate the matrix-vector product $\mathcal J\bm s_k$ at every $k$-th iteration with $\mathcal O(\mathcal N^2)$ operations, where $\mathcal J$ is the Jacobian matrix and $s_k$ is the error to be found. In order to reduce this high cost to $\mathcal O(\mathcal N)$, let us consider the Frechet derivative
$$\mathcal J(\Psi_i^{k,n+\frac{1}{2}})s_k \approx \frac{\mathcal L(\Psi_i^{k,n+\frac{1}{2}}+\eta s_k)-\mathcal L(\Psi_i^{k,n+\frac{1}{2}})}{\eta},$$ 
where $\eta$ is a small real number. The performance of this approach -- known as the Jacobian free Newton-Krylov~(JFNK) algorithm -- was verified for multiphysics simulations. However, the improvement in the operation count is paid off by requiring a preconditioner, which is a serious drawback for extending the JFNK to the simulation of heat and mass transfer. In the present work, we study an alternative, where the multiscale wavelet method captures the multiscale physics, and the Newton's method along with the Frechet derivative is used to reduce the residual of~(\ref{eq:bvp}) by a certain fraction at each level of the present multiscale wavelet based solution methodology.  
\subsection{The multiscale wavelet methodology}
The wavelet method~(see~\cite{Alam2012,Mallat,Sweldens97}) captures the multiscale physics, using the best $\mathcal N$ terms of the multiscale decomposition 
\begin{equation}
  \label{eq:awcm}
  \Psi_j(x_i) = \sum_{k}c_{2k}^s\varphi_{2 k}^s(x_i)
  +
  \sum_{l=s}^{\infty}\sum_{\eta=0}^{2^d-1}\sum_{k}^{}d_{2k+1}^{\eta,l}\psi_{2 k+1}^{\eta,l}(x_i),
\end{equation}
according to a prescribed error tolerance $\epsilon$ on the magnitude of $d_{2k+1}^{\eta,s}$, where the error for resolving mass and/or energy is $\mathcal O(\epsilon)$. In~(\ref{eq:awcm}), the first term represents a sampling of $\Psi_j$~($u_1$, $u_3$, or $c$) on a coarse grid of $(2^s n+1)\times (2^s m+1)$ points, and the second term models the additional details of $\Psi_j$, which is not captured by the first term. In $2$D, with the exception of east and north boundaries, each coarse grid sampling $c_{2k}^s$ corresponds to three details data $d_{2k+1}^{\eta,s}$~(for $\eta=0,\,1,\,2$). For example, if the REV $[0,4]\times[0,2]$ is divided by a factor of $2$ in each direction, for the $2k\,$th vertex $(0,0)$, we have three $(2k+1)$ neighbors $(0,1)$, $(2,0)$, and $(2,1)$, where the corresponding detailed information is stored. When a REV is not sufficient to resolve $\Psi_j$ on the $2k$-th grid point, at least one $d_{2k+1}^{\eta,s}$ will have a magnitude larger than $\epsilon$, thereby providing with a natural framework for adapting the size of the REV to the local physical property.  Therefore, one would begin with REVs on a coarse grid, perform a wavelet analysis, and recursively resize only those REVs, where wavelet coefficients are large. The wavelet transform may be computed with the Wavelet toolbox of Matlab without knowing the explicit information of $\varphi_{2k}^s$ and $\psi_{2k+1}^{\eta,s}$. Note that the grid adaptation is automatic with the wavelet method~\cite{Oleg2005}.    %

\begin{figure}
   \centering
    \includegraphics[trim=0cm 0cm 0cm 0cm,clip=true,width=9cm,angle=00]{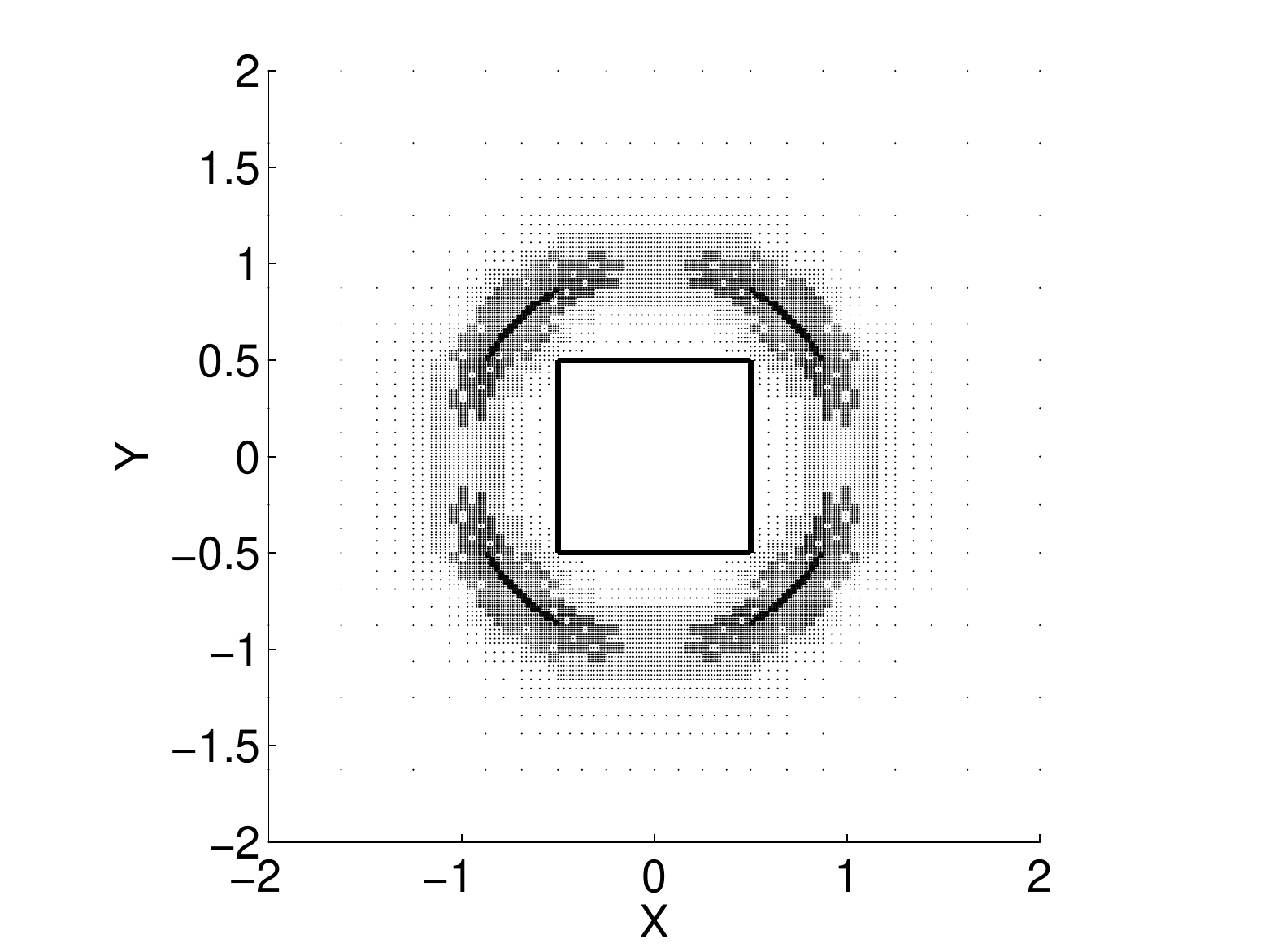}
   \caption{An example of the wavelet based adaptive grid generation for a typical concentration field $c(x_1,x_3)$, where the 'box' at the center of the domain represents an impermeable region. Each 'dot' represents the physical position of a wavelet, representing a REV. More wavelets are used near a circle of radius $1$, where $c(x_1,x_3)$ has a sharp change, which confirms that REVs have been adapted to capture the local physical variation.}
   \label{fig:cylg}
 \end{figure}

To illustrate the benefits of the multiscale wavelet representation, consider a prescribed concentration $c(x_1,x_3)$ that decays exponentially with respect to a circle of radius $1$. Taking the wavelet transform of $c(x_1,x_3)$, and recursively adapting the grid until all new wavelet coefficients satisfy a tolerance $\epsilon = 10^{-4}$, an adaptive wavelet grid has been obtained, which is presented in Fig~\ref{fig:cylg}. Note that an obstacle of size $[-0.5,0.5]\times[-0.5,0.5]$ has been placed at the center of the domain, showing that the wavelet transform can be computed on complex domains. 

\section{Results and verification}\label{sec:ver}
\subsection{Verification results}
\begin{figure}
  \centering
  \includegraphics[trim=0cm 4cm 0cm 0cm,clip=true,width=9cm,angle=00]{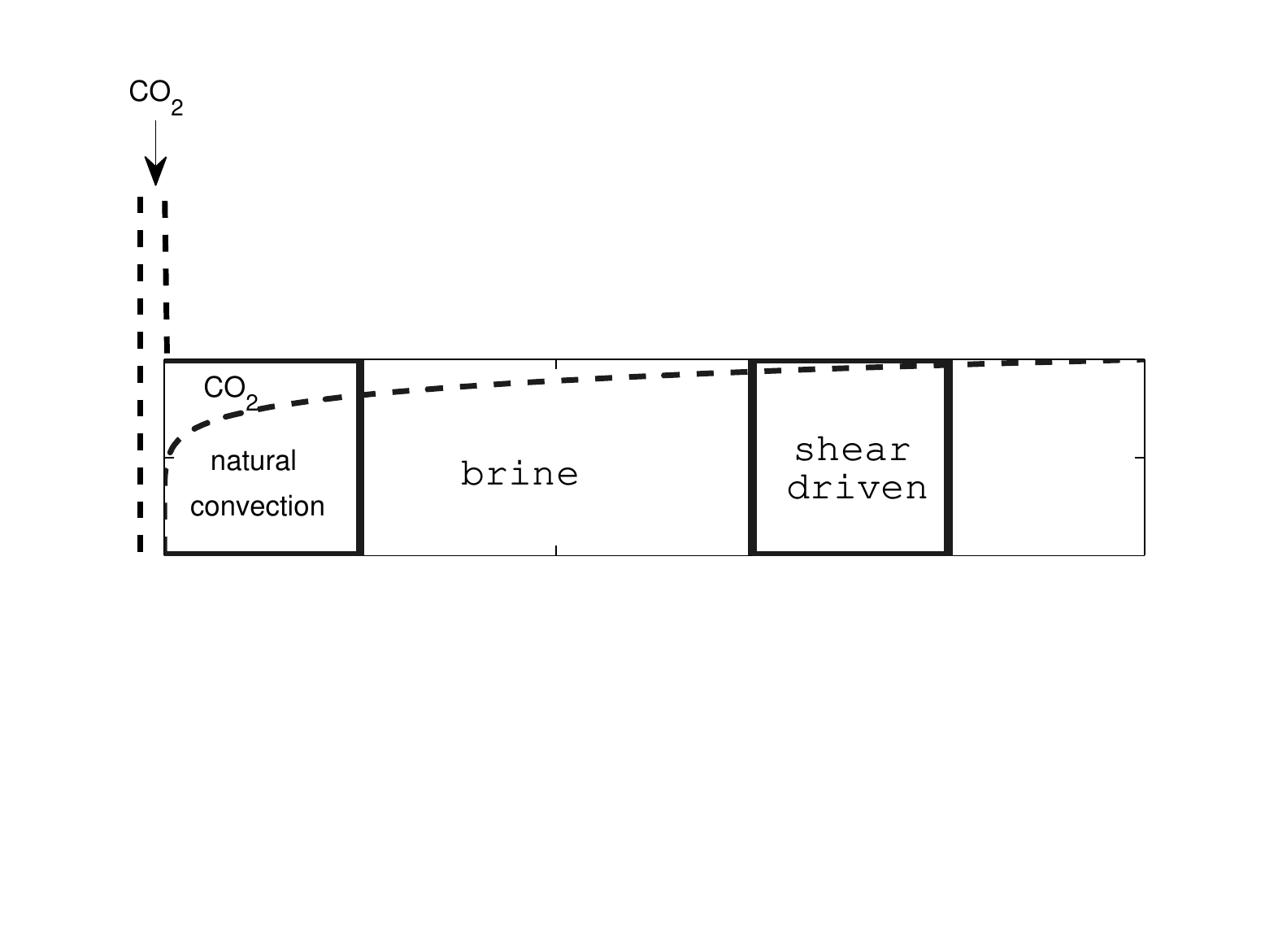}
  \caption{Schematic description for the distribution of the CO$_2$ phase after it has been injected into an aquifer. The natural convection mass transfer is an idealized model in the region that is next to the injection well - marked by the box. A region is marked ``shear driven'', which may be a typical high permeability zone. Note that a shear driven case has been considered for the purpose of numerical verification only.}
  \label{fig:ideal}
\end{figure}

The shear driven or natural convection transport of CO$_2$ in an aquifer is an idealized model, \Add{and is useful for computational verification}, where CO$_2$ moves horizontally just below the impermeable caprock, or vertically after it has been injected through an injection well. Adapted from~\cite{Pruess2008,Nordbotten2006}, a shear driven case and a natural convection case have been shown schematically in Fig~\ref{fig:ideal}. Note that the shear driven case has been chosen for the availability of reference data \Add{so that the numerical model can be quantified}. In the next two simulations, the parameters $\mathcal Da = 10^4$, $\mathcal Re=10^3$, $\alpha=1$, $\mathcal Sc=0.72$, $\phi=90\%$ have been adapted from~\cite{Yang2012}. 

\subsubsection{A shear driven mass transfer model for $\mathcal Gr\rightarrow 0$}
The shear driven mass transfer at $\mathcal Gr=0$ is a benchmark example, where the momentum transfer is independent of the mass transfer.
To study the performance of the proposed multiscale model, results of $30$ simulations of a shear driven mass transfer case with $\mathcal Gr=0$  have now been summarized, where for each of the CFL numbers are~$1$, $2$, $3$, $4$, $5$, $6$,  the tolerance values are $\epsilon=10^{-6}$, $10^{-5}$, $10^{-4}$, $10^{-3}$,~$10^{-2}$. In the present multiscale model, $\epsilon$ controls $\mathcal N$, as well as $\Delta x=\min(\Delta x_1, \Delta x_3)$ -- the length scale of the smallest REV, where the grid points are adapted dynamically. To utilize the full advantage of this adaptivity, the maximum time step, $\Delta t$, has also been adapted dynamically based on CFL~=~$\Delta t/\Delta x$. Clearly, an increase of the CFL number (or $\epsilon$) would increase the global temporal (or spatial) error. These $30$ simulations provide an understanding of how the spatial and temporal errors are controlled globally in the present model. Note the use of a large CFL number (CFL=$6$) is a distinct feature of the present methodology, and is an important contribution to the field of computational heat and mass transfer.

\begin{figure}[htbp]
  \centering
  \begin{tabular}{cc}
    \includegraphics[trim=1.75cm 0cm 0cm 0cm,clip=true,width=6cm,angle=00]{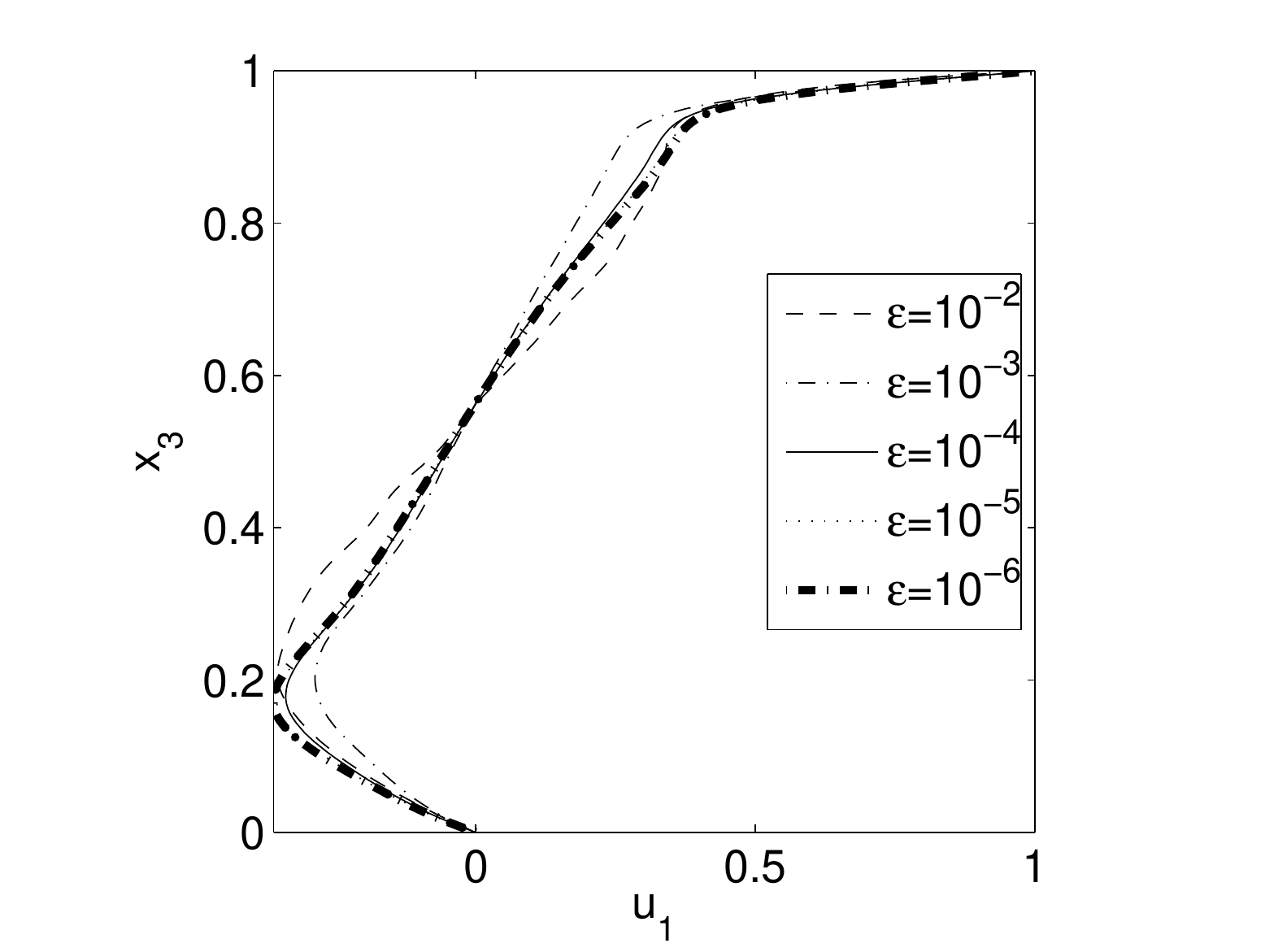}
    &
    \includegraphics[trim=1.75cm 0cm 0cm 0cm,clip=true,width=6cm,angle=00]{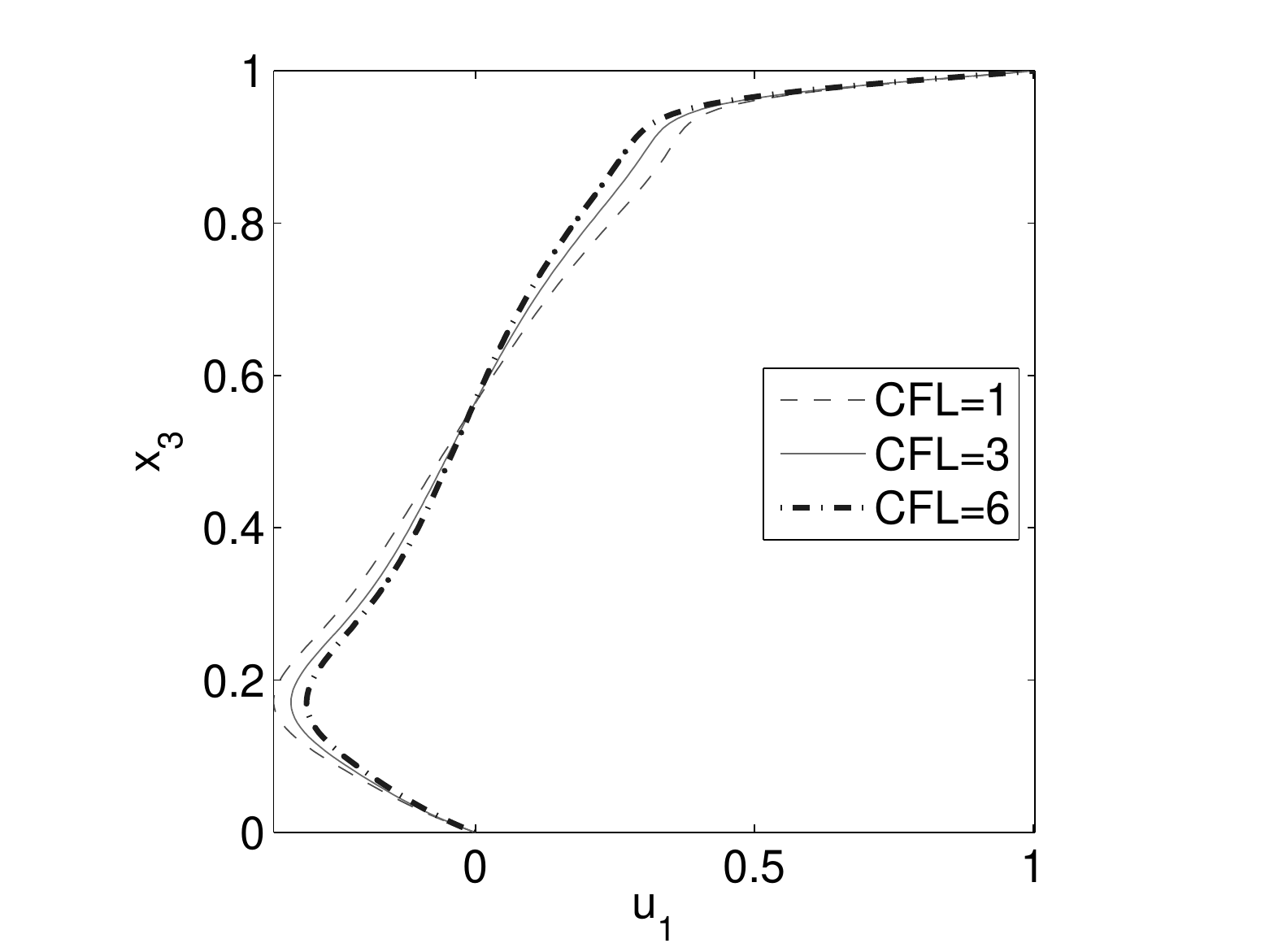}
    \\
    $(a)$ & $(b)$\\
    \psfrag{N}[c]{$\mathcal N$}
    \includegraphics[trim=1.75cm 0cm 0cm 0cm,clip=true,width=6cm,angle=00]{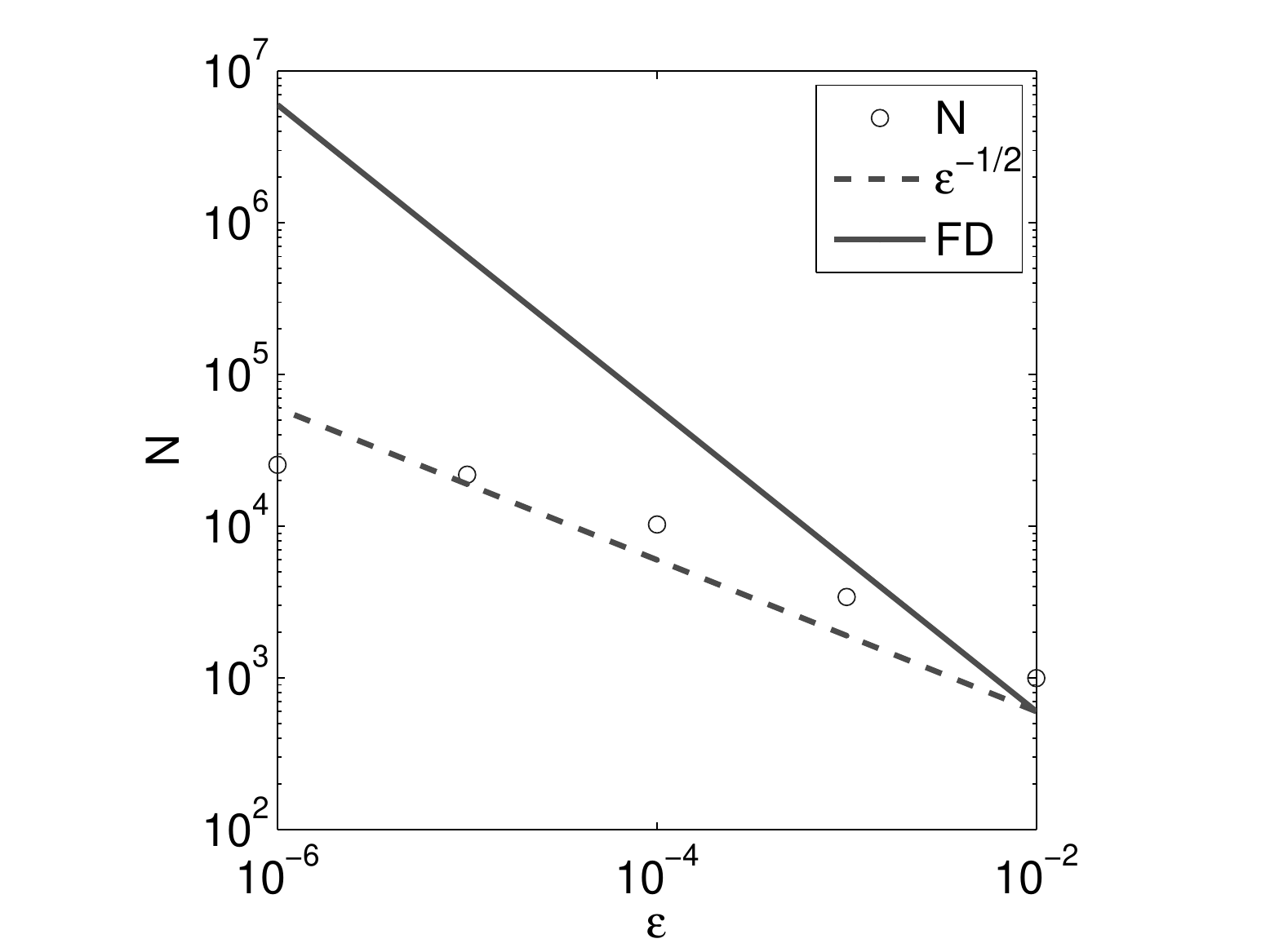}
    &
    \psfrag{Re}[c]{$\mathcal Re$}
    \includegraphics[trim=1.6cm 0cm 0cm 0cm,clip=true,width=6cm,angle=00]{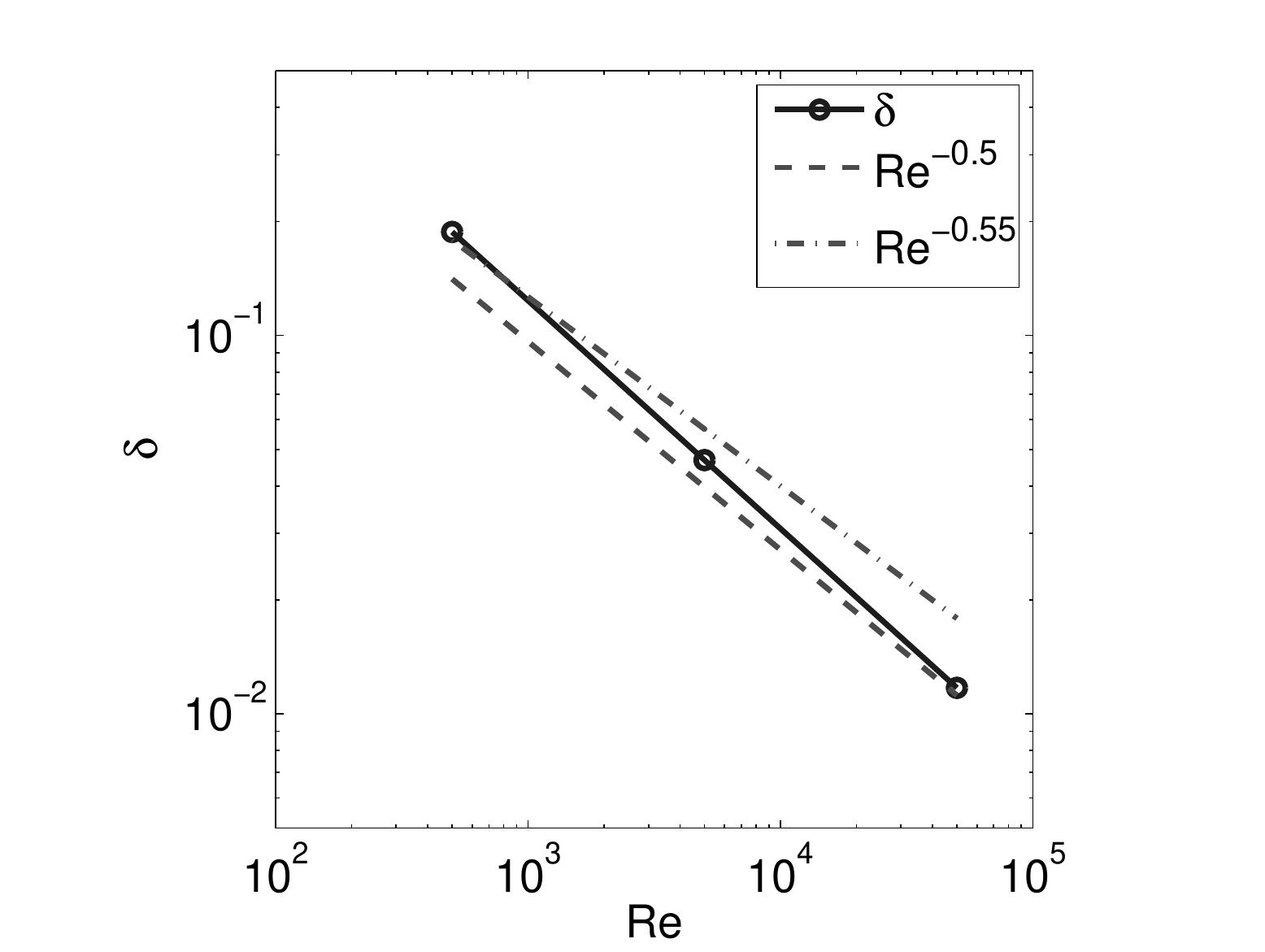}
    \\
    $(c)$ & $(d)$\\
  \end{tabular}
  \caption{Results from the performance studies of the proposed multiscale mode. The horizontal velocity $u=u_1\times\phi/U~\nicefrac{m}{s}$ as a function of $z=x_3\times H~\unit{m}$ along the vertical centerline of the aquifer. $(a)$~CFL~$=3$ is fixed, but the tolerance, $\epsilon$, varies in the range $10^{-6}\le\epsilon\le 10^{-2}$. $(b)$~The tolerance $\epsilon=10^{-4}$ is fixed, but the CFL number varies in the range, $1\le\hbox{CFL}\le 6$. Clearly, the present model controls the space-time variation of the error, which is an important objective of this development. $(c)$~The number of degrees of freedom, $\mathcal N$, as a function of tolerance, $\epsilon$. The result for the present model, $\mathcal N\sim\epsilon^{-1/2}$, is compared with that of a classical finite difference~(FD) model, $\mathcal N\sim\epsilon^{-1}$. $(d)$~Estimated boundary layer width, $\delta$, has been compared with $\mathcal Re^{-0.5}$ and $\mathcal Re^{-0.6}$.}
  \label{fig:shear}
\end{figure}

As demonstrated in Fig~\ref{fig:shear}($a$-$b$), the variation of the spatial error tolerance $10^{-6}\le\epsilon\le 10^{-2}$ and the CFL number, $1\le\hbox{CFL}\le 6$ confirms the space-time error control of the proposed model. This velocity profile has a good agreement with the one, which was presented by~\citet{Yang2012} and~\citet{Guo2002}. Fig~\ref{fig:shear}($c$) shows that the computational degrees of freedom, $\mathcal N$, varies like $\epsilon^{-1/2}$, which means that $\mathcal N$ does not increase linearly if $\epsilon$ is decreased. Note that a direct numerical simulation of heat and mass transfer phenomena with a classical finite difference~(FD) model, would increase $\mathcal N$ linear with a reduction of the error. The efficiency of the present model can be seen from the comparison in Fig~\ref{fig:shear}($c$). Further more, with the error tolerance, $\epsilon=10^{-3}$, the present model needs $\mathcal N=3\,416$, which is about $5\%$ and $13\%$ of the grid points required by the simulations of~\citet{Ghia82} and~\citet{Botella98}, respectively. Although the physical setting of the present simulation is quite different than that of refs~\cite{Ghia82,Botella98}, the similarity of dimensionless parameters allows us for a brief comparison in order to provide with a hint to the potential of the present model.   

The laminar boundary layer thickness just below the top of the aquifer has been compared with the theoretical estimate, $\delta\sim\mathcal Re^{-1/2}$, in Fig~\ref{fig:shear}($d$). A good fit between $\mathcal Re^{-0.5}$ and  $\mathcal Re^{-0.6}$ indicates that the laminar boundary layer has been resolved more accurately at higher $\mathcal Re$. This idealized simulation at $\mathcal Gr=0$ exploit the potential of the present model for simulating the best solution using the least number of the degrees of freedom, $\mathcal N$.

\subsubsection{A natural convection case for $\mathcal Gr\rightarrow\infty$}
\begin{figure}[htbp]
  \centering
  \begin{tabular}{cc}
    \psfrag{N}[c]{$\mathcal N$}
    \psfrag{Gr}[c]{$\mathcal Gr$}
    \includegraphics[trim=1.75cm 0cm 2cm 0cm,clip=true,width=6cm,angle=00]{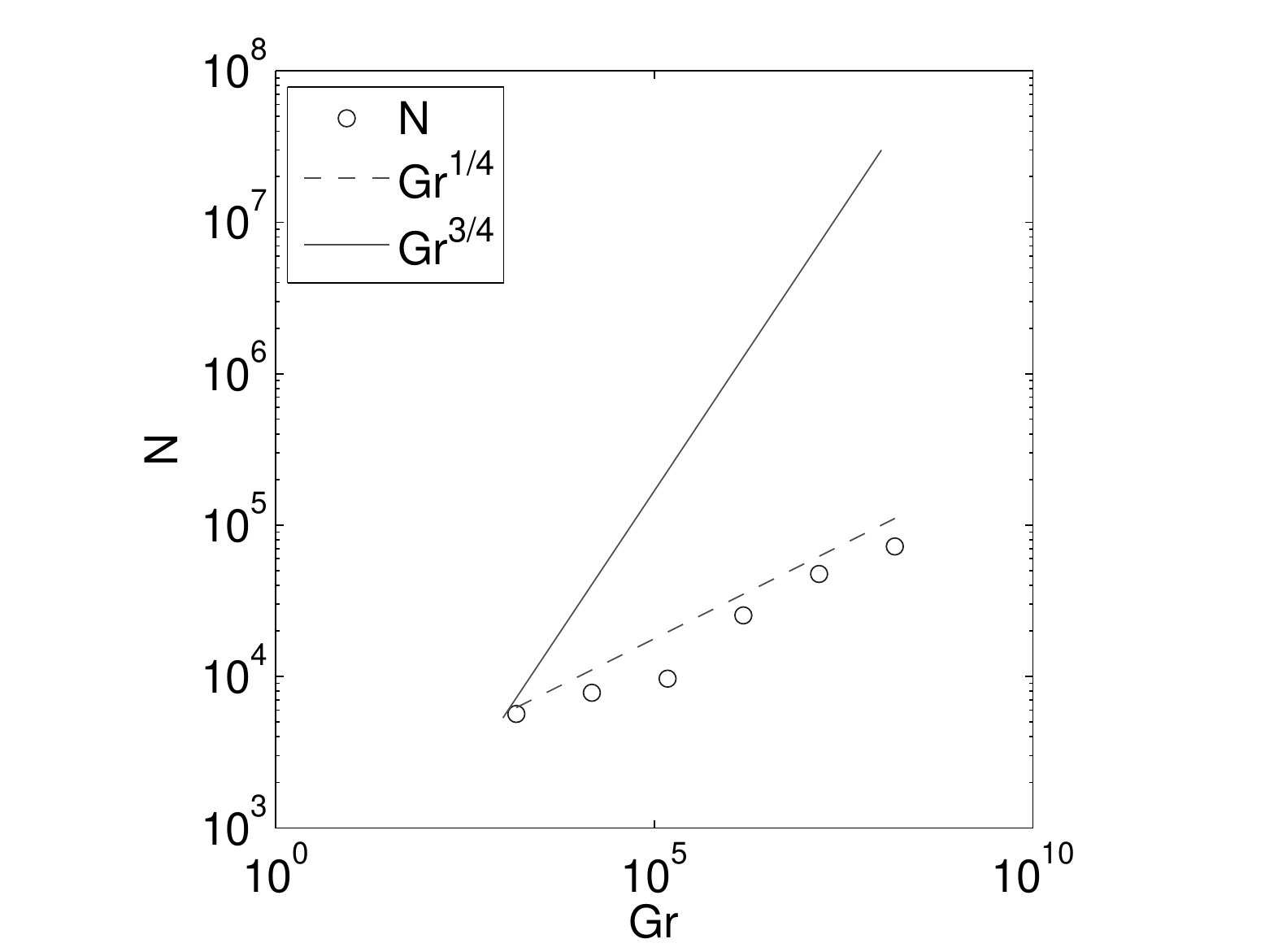}
    &    
    \psfrag{N}[c]{$\mathcal N$}
    \psfrag{Gr}[c]{$\mathcal Gr$}
    \includegraphics[trim=1.75cm 0cm 2cm 0cm,clip=true,width=6cm,angle=00]{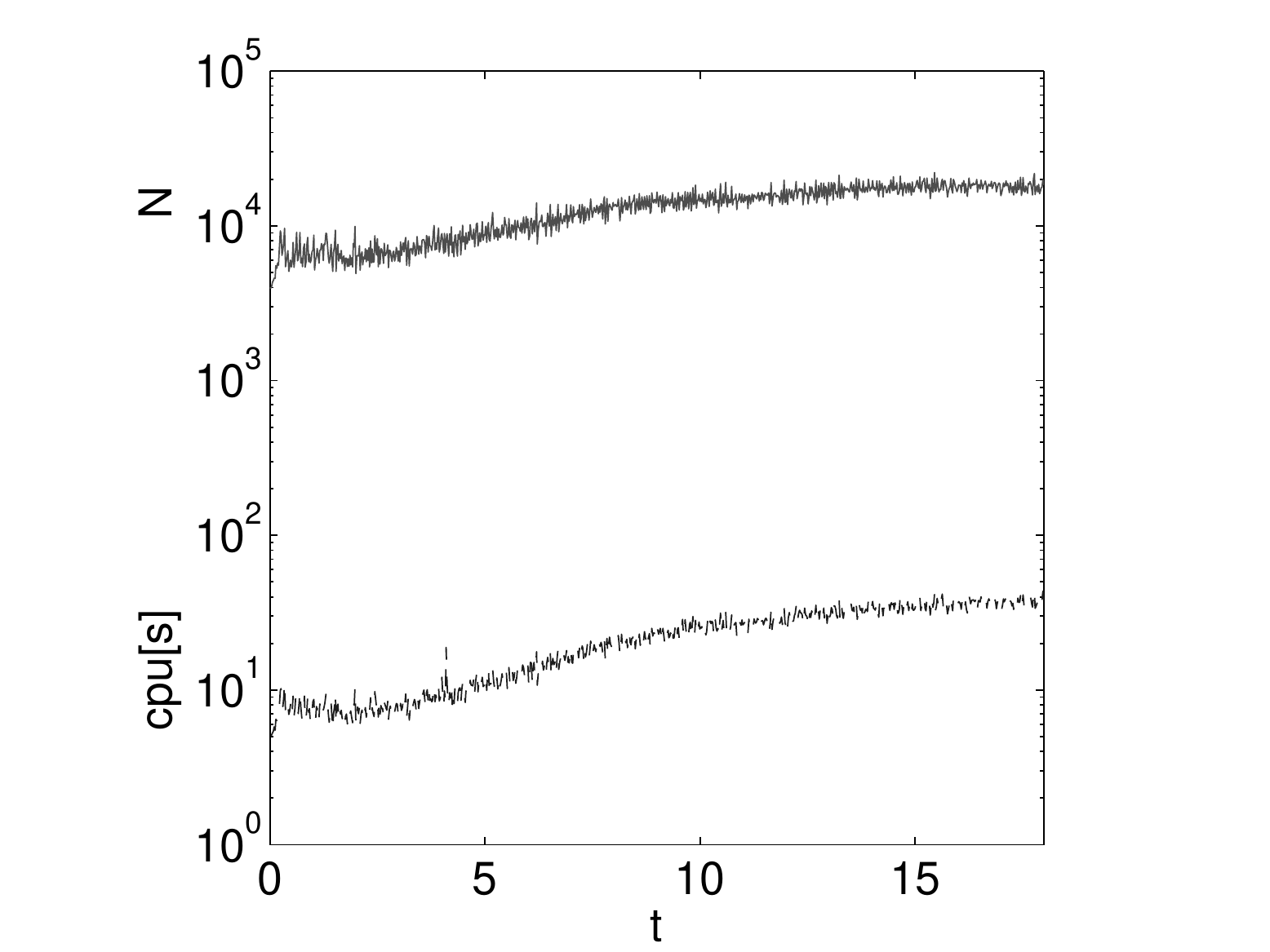}
    \\
    $(a)$ & $(b)$
  \end{tabular}
  \caption{$(a)$~The number of degrees of freedom ($\mathcal N$) for resolving the multiscale mass transfer as a function of $\mathcal Gr$ has been compared between the present multiscale model and a classical mass transfer model. The computational gain of a multiscale model at high $\mathcal Gr$ is evident. $(b)$~The scaling of the CPU time with respect to $\mathcal N$, which shows that the CPU time remains approximately proportional to $\mathcal N$ for the entire period of simulation.}
  \label{fig:GrN}
\end{figure}

An idealization of the natural convection mass transfer in the aquifer has been simulated, where the boundary conditions, $C=1$ and $C=0$ on $x_1=0$ and $x_1=1$, \Add{represents} the injection and removal of CO$_2$ across $x_1=0$ and $x_1=1$, respectively. Other boundary conditions are $\frac{\partial C}{\partial x_3}=0$ on $x_3=0$ and $x_3=1$, $\frac{\partial u_1}{\partial x_1}=0$ on $x_1=0$ and $x_1=1$. The velocity is assumed zero on all other boundaries. Under similar conditions, \citet{Nordbotten2006} studied mixed convection mass transfer in an aquifer, and showed that the migration of CO$_2$ occurs beneath the top boundary of the aquifer when natural convection dominates over the forced convection. Clearly, at large $\mathcal Gr~(\rightarrow\infty)$, when the natural convection is dominant, the numerical simulation of the mass transfer in an aquifer would require high spatial resolution in order to resolve the solutal boundary layer. Hence, an estimate for $\mathcal N$ as a function of $\mathcal Gr$ provides with a good understanding for the efficiency of the present multiscale model. Here, we summarize $7$ representative simulations with values of $\mathcal Gr$ ($1.5\times 10^3$, $1.5\times 10^4$, $1.5\times 10^5$, $1.5\times 10^6$, $1.5\times 10^7$, $1.5\times 10^8$, and $1.5\times 10^9$), which show that $\mathcal N$ increases approximately as $\mathcal Gr^{1/4}$. In  Fig~\ref{fig:GrN}, this result is also compared with the estimate $\mathcal Gr^{3/4}$ from the classical numerical simulation.  

\subsection{Simulation of a vertically migrating CO$_2$ plume}
\subsubsection{The Carrizo-Wilcox aquifer in Texas~(CWT) }\label{sec:CWT}
Over a period of $50$~years, the injection of approximately $370\times 10^6~\unit{m}^3$ super-critical CO$_2$ per year into the central section of the CWT would store about one fifth of the CO$_2$ emissions in Texas~\cite{Nicot2008,Hesse2008}.  The CWT formation is about $200~\unit{m}$ deep, and extends for a length of about $110~\unit{km}$ at an angle~$\approx 1.5^{\hbox{\tiny{o}}}$ with the horizontal, reaching a depth of about $4~\unit{km}$ beneath the earth's surface. 
A 2D vertical section of the CWT beneath a horizontal caprock at a depth of $3~\unit{km}$, which is $200~\unit{m}$ thick, and $2~\unit{km}$ long, has been simulated. The length to width aspect ratio of the model region is $10:1$, which is different than the actual aspect ratio of the CWT. The temperature of the aquifer is assumed at the initial value. A $400~\unit{m}$ wide CO$_2$ plume is assumed instantaneously at the center of the bottom boundary. The migration of CO$_2$ takes place under the action of gravity.
Initially, both the resident brine and the invaded CO$_2$ are assumed at rest. The normal components of the gradient of mass and momentum fluxes are assumed zero on the lateral boundaries for all time.

\begin{figure}
  \centering
  \includegraphics[height=8cm]{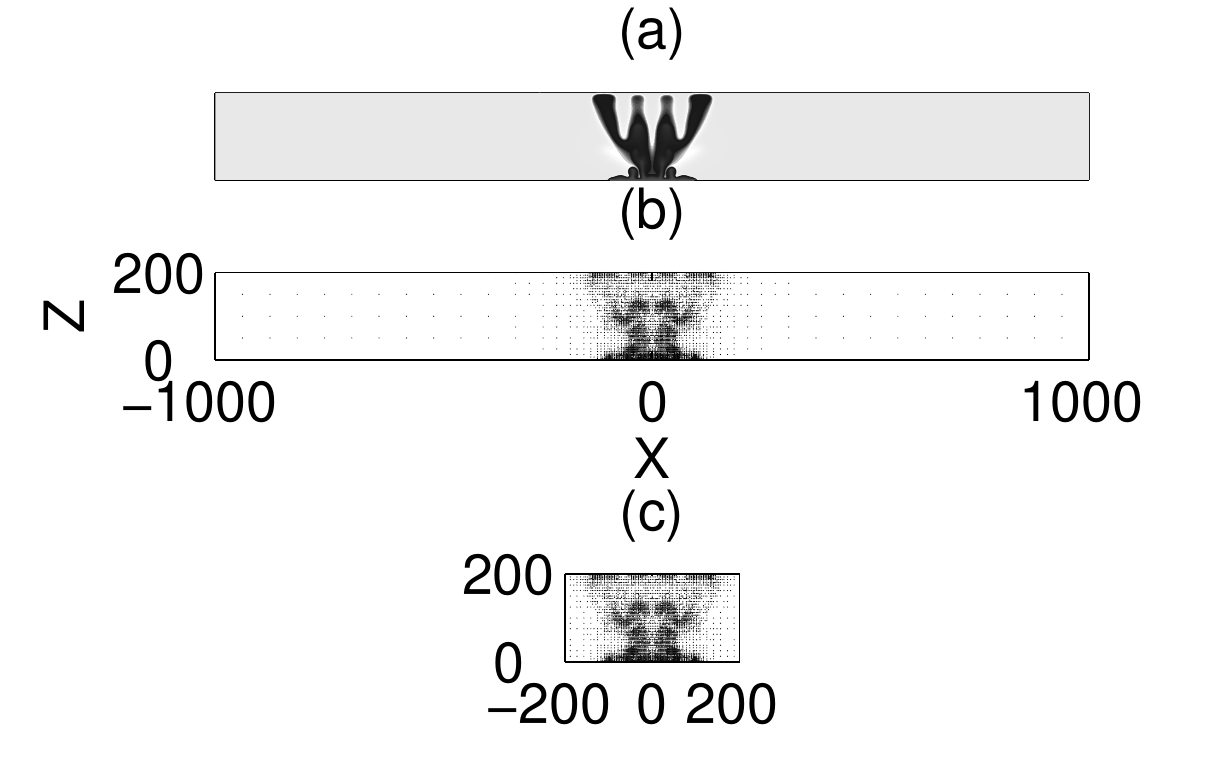}
  \caption{Simulated plume migration and associated fingers in a $2,000~\unit{m}\times 200~\unit{m}$ domain. $(a)$~Distribution of CO$_2$ is marked with the dark color, where the light color represents brine. $(b)$~Distribution of grid points, showing that the most significant flow is located only in the region of fingers. $(c)$~The most significant flow occupies a part $400~\unit{m}\times 200~\unit{m}$ of the domain, where a high resolution is needed.} 
  \label{fig:fing}
\end{figure}

In Fig~\ref{fig:fing}, simulated vertical migration of a CO$_2$ plume and the corresponding adapted grid points are presented at $t=11.4$ year. Clearly, the REVs have been adapted dynamically to resolve sharp gradients of the plume with respect to the error tolerance $\epsilon=10^{-4}$. The minimum size of a REV is given by $\Delta x_1\sim 2~\unit{m}$ and $\Delta x_3\sim 1.5~\unit{m}$ near the center of the plume, and the maximum size of the same is given by $\Delta x_1^H=62.5~\unit{m}$ and $\Delta x_3^H=50~\unit{m}$ away from the plume. The simulation needed an average $\mathcal N=3,593$ at each time step.

As depicted in Fig~\ref{fig:fing}(c), the plume remains approximately in a region that extends from $-200~\unit{m}$ to $200~\unit{m}$ in the horizontal direction with respect to the center of the domain. However, only in this part of the domain, one might get the qualitatively equivalent results by solving the multiscale model~(\ref{eq:inc2d}-\ref{eq:mnc}) with a classical numerical method, using the uniform resolution of $\Delta x_1\sim 2~\unit{m}$ and $\Delta x_3\sim 1.5~\unit{m}$. In that case, at least $26,000$ grid points would be needed. In comparison to this estimate, the present methodology (with $\mathcal N\sim 3,593$) is about $86\%$ efficient, albeit a direct comparison with a previous simulation would be realistic.

\subsubsection{A comparison of the computational gain with a {\rm TOUGH2} simulation}\label{sec:tough2}

The recorded CPU time for the present simulation is approximately $2~\unit{ms~to}$ $3.6~\unit{ms}$ per grid point on a Dell T7400 workstation. A useful reference on the computational effort that is needed for a similar simulation is given by~\citet{Pruess2011}, where the reported CPU time is about $700~\unit{ms}$ per grid block using the parallel code, TOUGH2, with $8$ processors in a Dell 5400 workstation. This  is equivalent to about $5\,600~\unit{ms}$ per grid block for a single processor computation. The CPU time comparison with this particular TOUGH2 simulation exploits the advantage of $\mathcal O(\mathcal N)$ computational complexity, which is one important development of the present multiscale model.

\begin{figure}
   \centering
   \begin{tabular}{cc}
   \includegraphics[height=8cm]{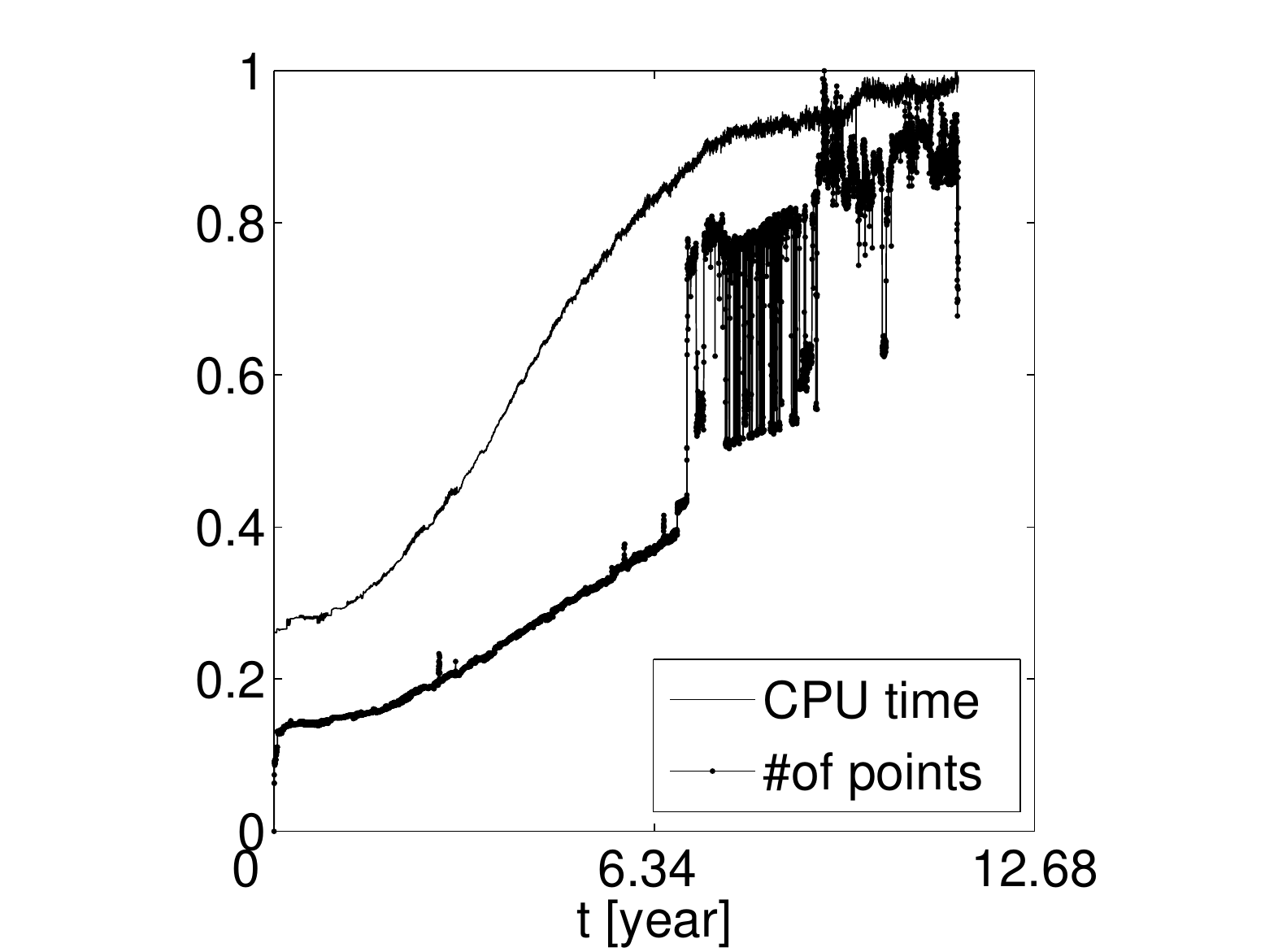}
%
   \end{tabular}
   \caption{A scaling of the CPU time is compared with the number of grid points. In order to fit both the curves in the same frame, both the CPU time (upper curve) and the number of grid points (lower curve) have been normalized with respect to the maximum value of the corresponding data. }
   \label{fig:d3cpu}
 \end{figure}

Since $\mathcal N$ varies at each time step due to adaptivity, we have recorded the CPU time for advancing the solution each time step, and two representative results are presented in Fig~\ref{fig:GrN}($b$) and Fig~\ref{fig:d3cpu}, showing that the CPU time is approximately proportional to $\mathcal N$. These results explore the promise of this development. 
\subsection{Analysis of natural convection mass transfer in an aquifer}

\subsubsection{Penetrative  mass transfer}
\citet{Ruck68} studied a generalized penetration theory for mass transfer in the vicinity of a fluid--fluid interface. In this section, we have briefly studied the multiscale nature of the penetrative mass transfer and the associated vorticity generation in a subsurface aquifer. In order to study the vorticity generation, we have simulated natural convection mass transfer at various values of $\phi$ and $\mathcal Da$.   

\begin{figure}
  \centering
  \begin{tabular}{c}
  \includegraphics[height=8.5cm]{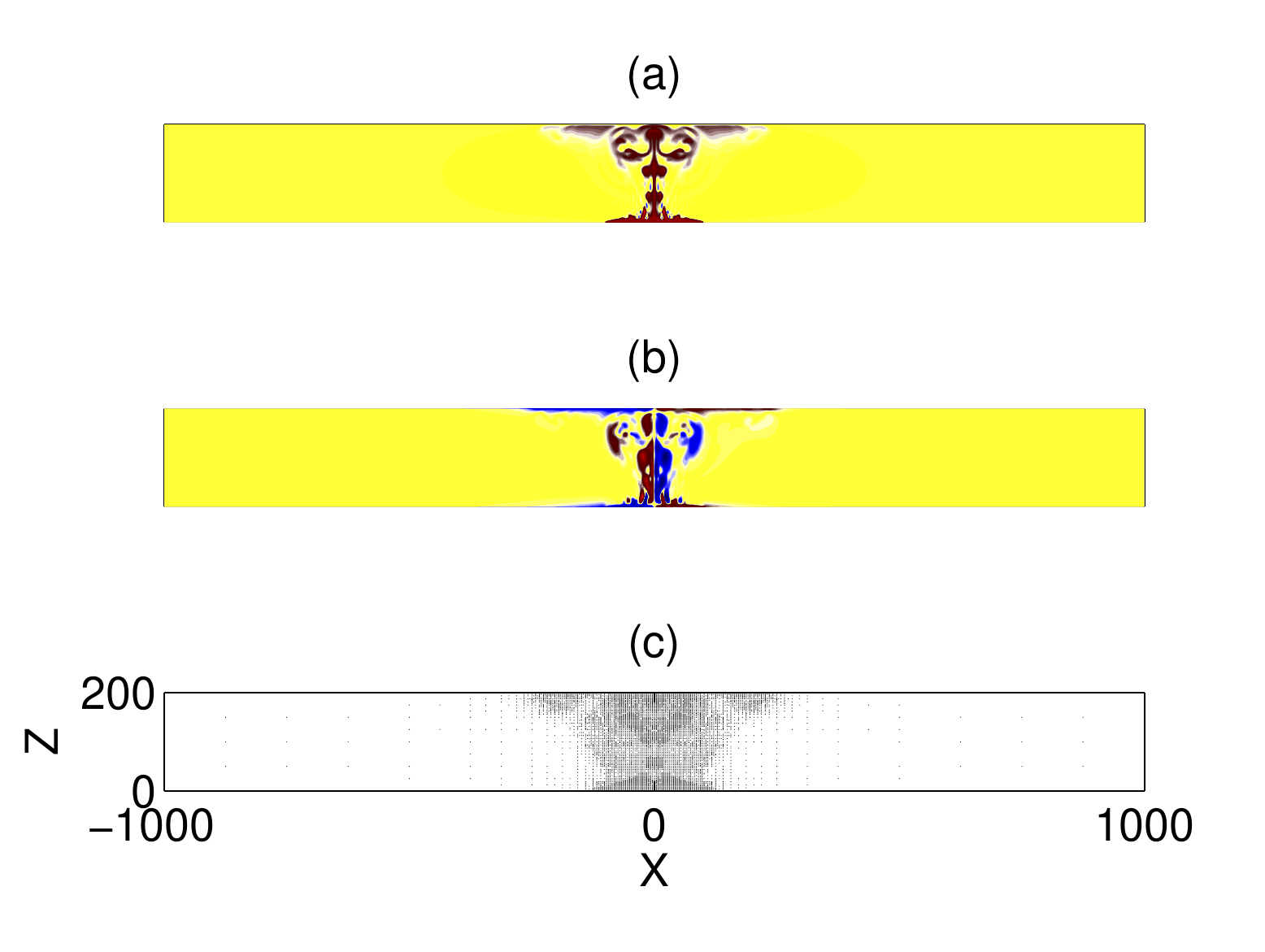}
  \\
  \includegraphics[height=8.5cm]{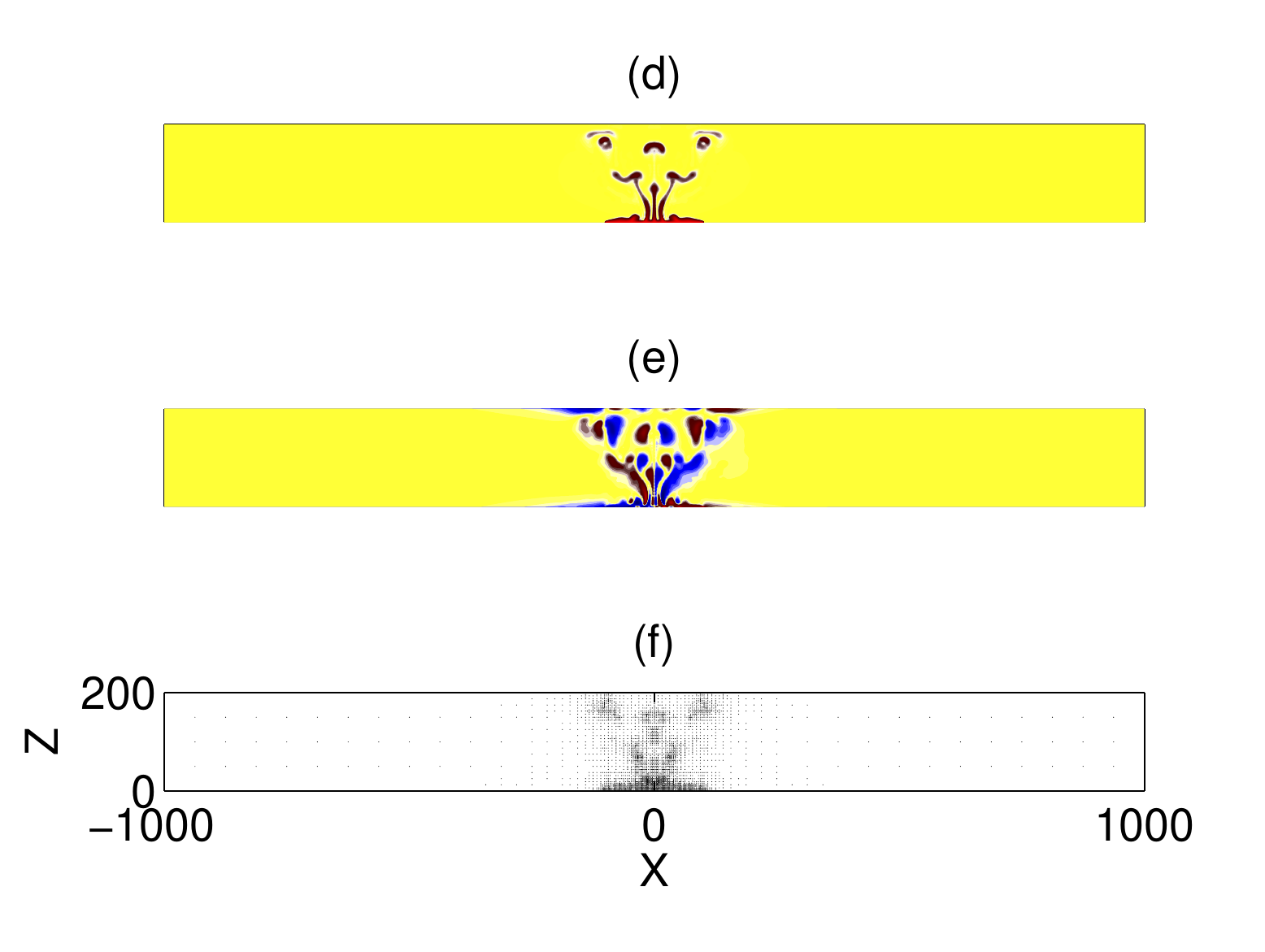}
  \\
  \end{tabular}
  \caption{The effect of the porosity ($\phi$) in a natural convection mass transfer and the associated vorticity generation. $(a,\,b,\,c)$ $\phi=10\%$, $(d,\,e,\,f)$ $\phi=20\%$; $(a,d)$~the concentration of CO$_2$, red: CO$_2$, yellow: brine; $(b,e)$~the vorticity $\omega_2$ associated with the natural convection plume, red: counter clockwise vortex, blue: clockwise vortex, yellow: zero vorticity; $(c,f)$~the grid points. In $(c)$, $\mathcal N=7,013$, and in $(f)$, $\mathcal N=3,553$. Clearly, when $\phi$ increases the interaction between the fluid and the porous structure weakens.}
  \label{fig:v3c}
\end{figure}

In the limit of $\mathcal Da\rightarrow 0$, a first order estimate for the vorticity as a function of concentration is $\omega_2\sim\frac{\partial c}{\partial x_1}$, which indicates that a counter clockwise vorticity is generated in the region that is to the immediate left of the plume when the plume migrates upward~\cite{Riaz2006}. Similarly, a clockwise vorticity is expected in the region that is to the immediate right of the plume.  These clockwise and counter clockwise vortices interact with the porous structure, accelerate horizontal migration, and are responsible for transferring mass and momentum from one scale to the other. In Fig~\ref{fig:v3c}, the plume, the vorticity, and the adapted grid are presented for the porosities, $\phi=10\%$ and $\phi=20\%$, where $\mathcal Da=2.7\times 10^{-4}$.  A complicated multiscale dynamics is observed. For the higher porosity, only $20\%$ of a REV is occupied by the fluid. A reduction of the porosity would decrease the volume of fluid faction in the REV, thereby strengthening the interaction between the fluid and the porous structure. For $\phi=10\%$, the mass transfer is accompanied with local sharp spatial gradients at multiple length scales, where $7,013$ grid points are needed because REVs are resized to resolve sharp gradients of the plume. If the porosity is doubled to $\phi=20\%$, the rate of mass transfer is enhanced by smoothing out the spatial gradients, and thus, the number of grid points is reduced to $3,553$. A better understanding of the vorticity field can be quantified with simulations at various values of $\mathcal Da$. %

\begin{figure}
  \centering
  \includegraphics[height=4.5cm]{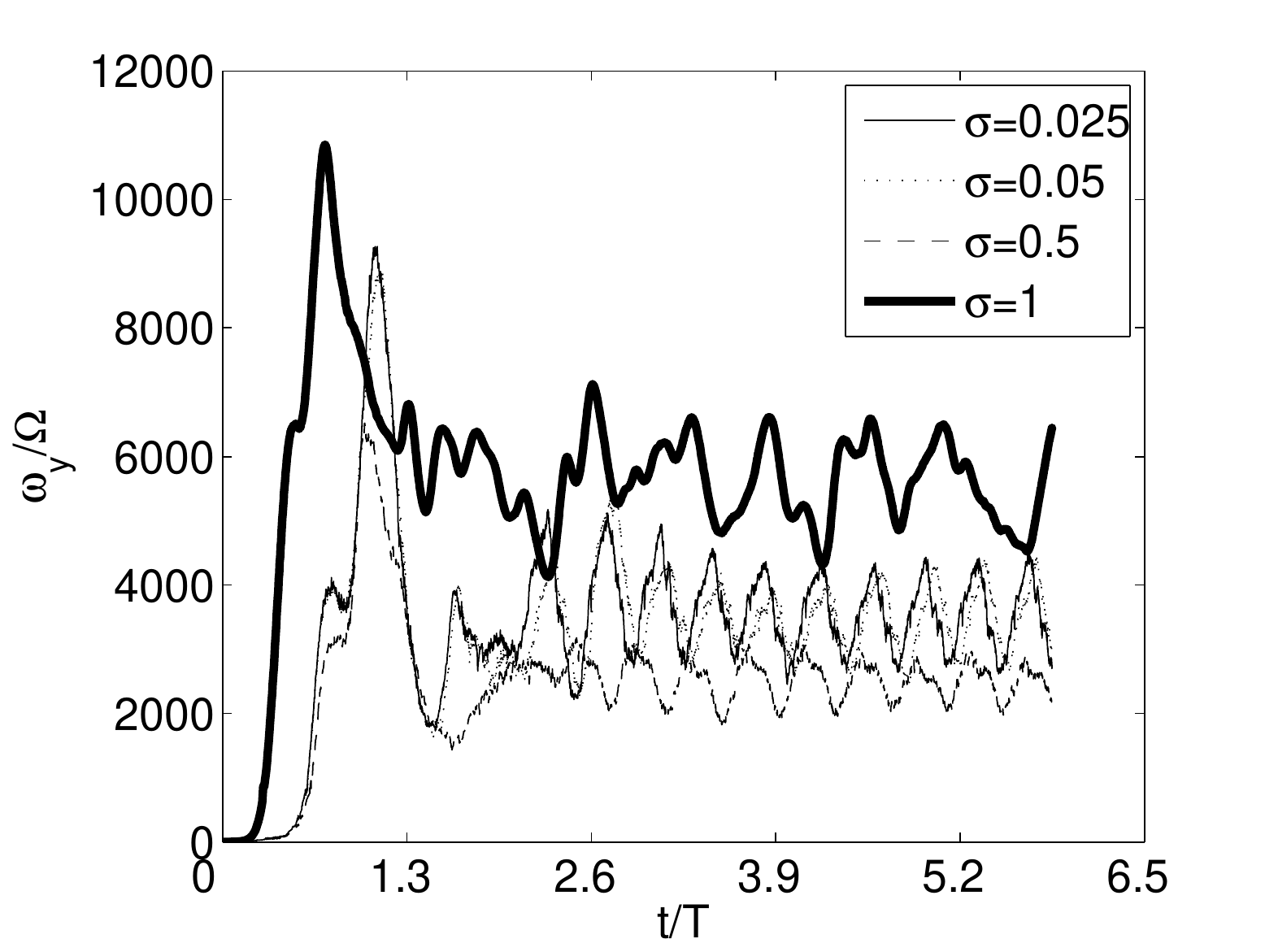}
  \caption{The time evolution of the mean vorticity $\frac{1}{A}\iint\omega_2(x_1,x_3,t)\,dA$, where $\omega_2 = \frac{\partial u_3}{\partial x_1}-\frac{\partial u_1}{\partial x_3}$  is presented for $\mathcal Da = 1.4\times 10^{-4},\,2.7\times 10^{-4},\,2.7\times 10^{-3},\,\hbox{and }5.4\times 10^{-3}$. The vorticity is normalized with respect to $\Omega = |\bm\nabla\times\bm u|$, and the time is normalized with respect to $T=L/U$, where $L$ and $U$ are length and velocity scales, respectively.}
  \label{fig:vordc}
\end{figure}

For the fixed $\phi=10\%$, let us present the time evolution for the mean vorticity for a variation of the Darcy number, $\mathcal Da=1.4\times 10^{-4},\, 2.7\times 10^{-4},\, 2.7\times 10^{-3},\,\hbox{and } 5.4\times 10^{-3}$. Since the variation of $\mathcal Da$ is between $10^{-4}$ and $10^{-3}$, these simulations explore the sensitivity of perturbations introduced by the porous structure into the penetrative mass transfer. The growth of the mean vorticity, and its time evolution has been presented in Fig~\ref{fig:vordc}. Clearly, the Darcy number affects both the maximum of the mean vorticity and its amplitude of the fluctuation. Note that the natural convection mass transfer enhances the vorticity at the lowest of these values of $\mathcal Da$, which indicates that the penetrative mass transfer in a saline aquifer is accompanied with vorticity.

\subsubsection{The effect of CO$_2$ dissolution into the natural convection mass transfer}\label{sec:Fr}
\begin{figure}
  \centering
  \begin{tabular}{c}
    $\mathcal Fr = \infty$\\
    \includegraphics[trim=1.5cm 3cm 1.5cm 3cm,clip=true,width=12cm]{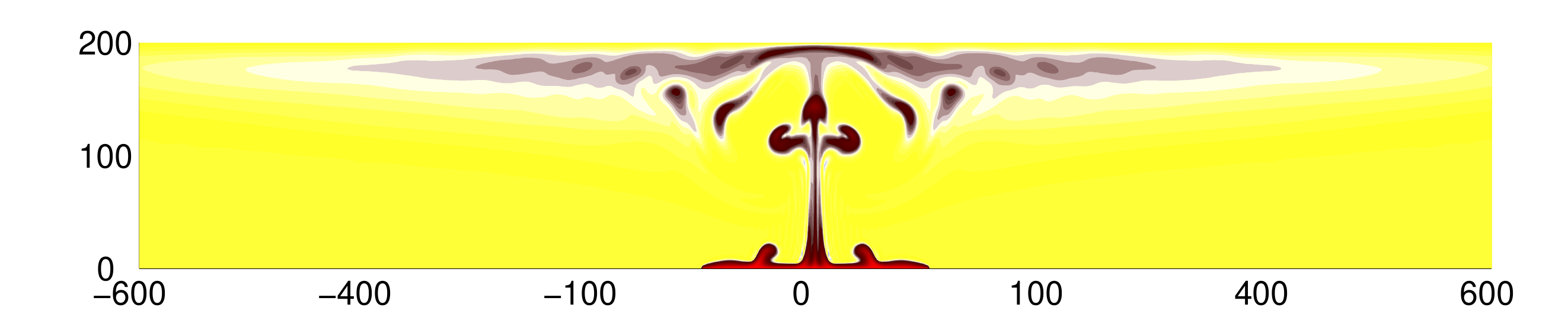}\\
    $\mathcal Fr = 3.2$\\
    \includegraphics[trim=1.5cm 3cm 1.5cm 3cm,clip=true,width=12cm]{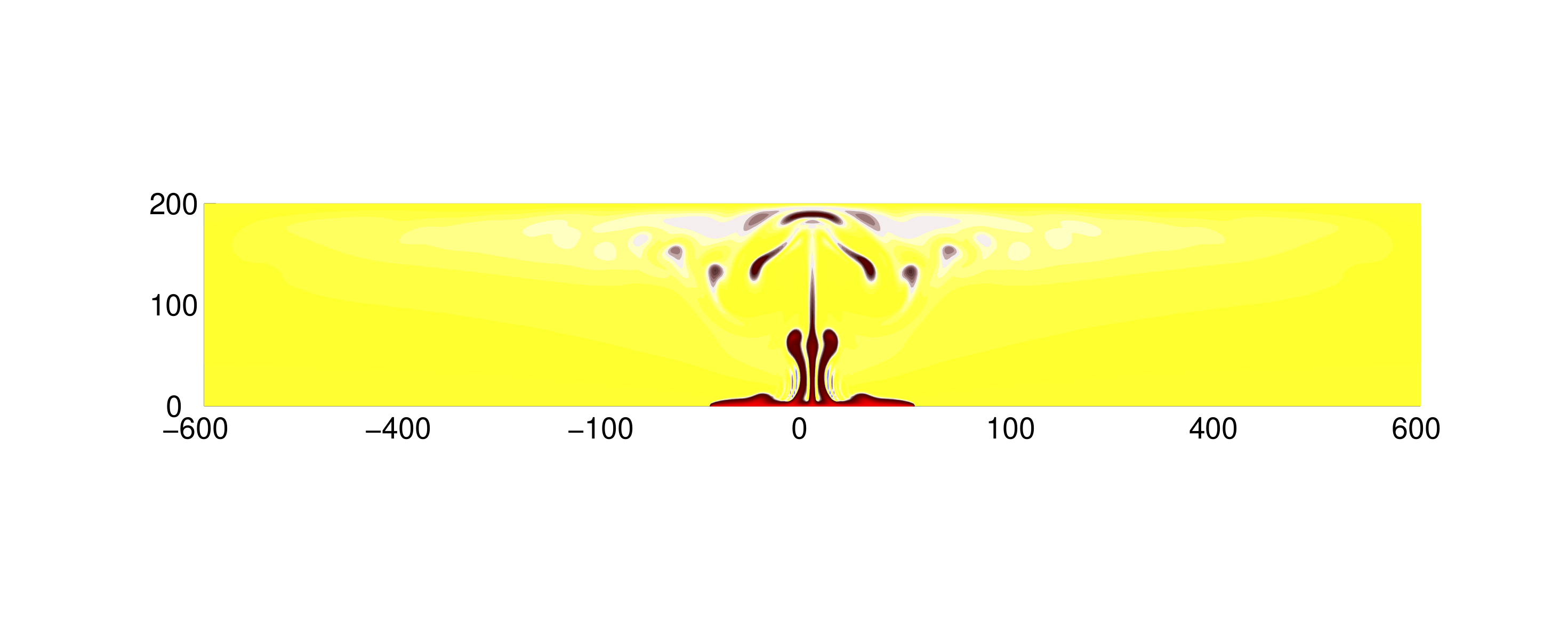}\\
    $\mathcal Fr = 1.0$\\
    \includegraphics[trim=1.5cm 3cm 1.5cm 3cm,clip=true,width=12cm]{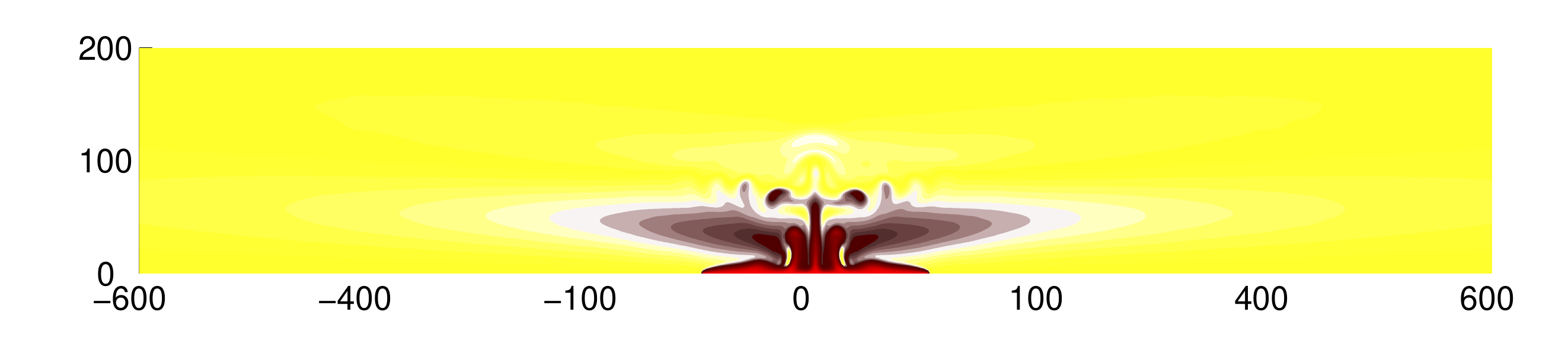}\\
  \end{tabular}
  \caption{Effect of dissolution on the vertical migration of a CO$_2$ plume for $\mathcal Fr=\infty,\, 3.2,\, 1.0$ at $t=11.4~\unit{year}$. The red and yellow represents $c=1$ and $c=0$, respectively. The displayed region extends from $-600~\unit{m}$ to $600~\unit{m}$ horizontally, and from $0~\unit{m}$ to $200~\unit{m}$ vertically.}
  \label{fig:CFR}
\end{figure}

In the present model, the dissolution of CO$_2$ into the resident saline has been characterized by the Froude number $\mathcal Fr$. 
In Fig~\ref{fig:CFR}, simulated distribution of CO$_2$ has been presented at $t=11.4~\unit{y}$ for three representative values of the Froude number, $\mathcal Fr=\infty$, $3.3$, and $1$. The plume migrates vertically until it reaches the impermeable caprock at the top of the reservoir when $\mathcal Fr=\infty$, when the brine is saturated with CO$_2$ such that $\Gamma=0$. When $1\le \mathcal Fr<\infty$, the plume migrates through a background environment with $\Gamma < 0$. Comparing Fig~\ref{fig:CFR}(a) with Fig~\ref{fig:CFR}(c), we see that the plume has traveled less than half way in the vertical direction when $\mathcal Fr=1$. 

This simulation indicates a marked variability in the vertical migration of CO$_2$. These two-dimensional idealized simulations hint on the potential of the present model for analysing natural convection mass transfer and other related phenomena associated with CO$_2$ storage in aquifers. Further analysis with three dimensional simulations would be useful for explaining the multiscale and complex phenomena associated with the carbon capture and storage program. 

\section{Summary and discussion}\label{sec:sum}
\subsection{Conclusion}
This article introduces a multiscale modeling and simulation approach for the natural convection mass transfer in an aquifer. In order to tackle the challenges of multiscale phenomena, we have adopted the classical volume averaging technique to model a full details of the multiscale transport. Taking average heat and mass transfer with respect to a REV leads to a general framework for parameterizing the effect of multiscale phenomena which is not resolved with the averaging process. A wavelet based multiscale simulation methodology has been studied to dynamically resize the REV so that localized mass transfer can be resolved efficiently. %
A brief summary of the present development and key findings have been outlined below.
\begin{itemize}
\item Using the space-time average of the first principle conservation laws with respect to a representative elementary volume and a representative elementary time scale, this work has studied the governing equations so that heat, mass, and momentum transfer can be captured in a range of multiple length and time scales. 
\item Results from a set of $30$ representative numerical experiments at $\mathcal Gr=0$ and $\mathcal Da=10^4$ show that the space-time error can be controlled with {\em a priori} prescribed error tolerance $\epsilon$ and CFL number. This {\em a priori} error control is an important contribution that would benefit future developments in the field of computational heat and mass transfer analysis. 
\item Simulations with an idealized natural convection mass transfer observes that the number of computational degrees of freedom varies like $\mathcal N\sim\mathcal Gr^{1/4}$. Clearly, exploiting multiscale physics in the computational model has the potential to compute the most significant proportion of the flow using a near optimal computational effort. Further more, all simulations have indicated that the CPU time remains approximately proportional to $\mathcal N$, which means that the amount of computational work remains proportional to the amount of actual physical change in the system. To the author's opinion, this property of the present model is a distinct feature with respect to classical approaches those are commonly used in the computational heat and mass transfer analysis.  
\item Present simulations with varying $\mathcal Fr$ show a marked variation in the vertical migration of a CO$_2$ plume, {\em i.e.} the associated time scale, which suggests that the effect of dissolution of CO$_2$ in saline has a dominant role on the mass transfer mechanism. This would help to explain the dynamics of the subsurface CO$_2$ plume in various storage facilities.
\item Present simulations observes that the vertical migration of a plume enhances the vorticity generation, where the dissolution of CO$_2$ has a strong influence on the vertical time scale of a plume.
\item To the best of knowledge, this article, for the first time, has extended the second generation wavelet based adaptive technique to the field of subsurface flow and transport modeling. Hence, the present works adds a novel technology to the growing interests of multiscale modeling in the field of computational transport in aquifers. 
\end{itemize}

\subsection{Discussions}
There are several possibilities to extend the present development. This includes the simulation of heat and mass transfer in a large aquifer -- such as $110~\unit{km}$ long CWT or any other similar sites -- where the phenomena is fully three-dimensional and gravitationally unstable. Future simulations, where the permeability, porosity, and other reservoir characteristics would have been obtained with advanced field technologies, may provide with an understanding of  the degree to the proportion of actual multiscale features that can be resolved by the proposed model within the constraint of modern powerful computing resources. In a more realistic $3$D simulation, if $\mathcal N$ is still constrained by the computer power, then the smallest scale of a REV may need to be increased, where further improvements of the parameterization scheme would be necessary. Therefore, the present work leaves potential open questions those may be addressed in the future. However, the present research indicates that $3$D simulation will be benefited  greatly if the present modeling approach is extended. Furthermore, the effect of variable permeability, \Add{temperature perturbation},  and pathway transmissivities have not been examined in the present article. These works are currently underway. 

\begin{appendices}
\section{}\label{app:ddm}
\section*{The space-time double decomposition methodology }
This section outlines the proposed model of resolving multiscale physics of a natural convection mass transfer when a plume migrates after CO$_2$ has been injected into an aquifer. The methodology aims to capture the average flow and transport with respect to a REV and a representative time scale for the REV, as well as to parameterize the effect of the unresolved flow. 
Researchers with interest in further mathematical details of the adopted double decomposition methodology may find the works of~\citet{DeLemos2006} and~\citet{Lage2002} useful. This section provides a brief outline for the decomposition of the inertia term $u_j\frac{\partial u_i}{\partial x_j}$ into a resolved proportion and an unresolved proportion, where the intrinsic space average ($\langle u_i\rangle$) is applied to time average ($\overline u_i$), and {\em vice-versa} (see also, section~\ref{sec:mimtm}). 

If we take the time average of a spatially averaged quantity, and use $\langle u''_i\rangle$ to denote temporal fluctuation of spatial averages, then a spatial mean may be decomposed as
$$
\langle u_i\rangle = \langle\overline u_i\rangle + \langle u''_i\rangle,
$$
where $\langle u''_i\rangle$ is the deviation of $\langle u_i\rangle$ with respect to the time average. The double decomposition is obtained by taking a spatial decomposition, $u_i = \langle u_i\rangle + \tilde u_i$, which is followed by a temporal decomposition; {\em i.e.,}
$$
u_i = \underbrace{\langle\overline u_i\rangle + \langle u''_i\rangle}
_{\langle u_i\rangle}
 +\overbrace{\overline u''_i + u''_i}^{\tilde u_i}
$$

The following assumptions have been adopted to parameterize the effect of unresolved multiscale physics.
\begin{itemize}
\item $\rho_0\overline{\langle u''_i\rangle\langle u''_j\rangle}$ corresponds to momentum flux perturbation associated with the temporal fluctuations of the spatially averaged quantities $\langle u''_i\rangle$, which has been neglected. This term needs to be parameterized if a transition to turbulence is important; {\em i.e.}, if $\mathcal Re_p\gg 1$.

\item $\rho_0\langle\overline u''_i\overline u''_j\rangle$ corresponds to momentum diffusion associated with the temporal mean of spatial fluctuation (same as the spatial average of temporal fluctuation). In order to allow a smooth transition of flow and transport through pores of a porous medium, \citet{Brinkman49} suggested to incorporate the effect of momentum diffusion. In the present work, this term has been parameterized to incorporate the effect of momentum diffusion.

\item $\rho_0\overline{\langle u''_iu''_j\rangle}$ represents dispersion of momentum due to both time and spatial fluctuation, and can be neglected unless turbulence intermittency is important.
\end{itemize}
The above approximations, along with the conservation of mass and the Boussinesq approximation, leads to $u_j\frac{\partial u_i}{\partial x_j} = \frac{\partial u_iu_j}{\partial x_j}$, and hence, the spatio-temporal average of the divergence of the momentum flux (divided by the density) can be derived recursively. First, let us apply the spatial decomposition, $u_i = \langle u_i\rangle + \tilde u_i$, and write
$$
\frac{\partial}{\partial x_j}\overline{\langle u_iu_j\rangle} =\frac{\partial}{\partial x_j}\overline{[\langle u_i\rangle\langle u_j\rangle + \langle\tilde u_i\tilde u_j\rangle]} 
$$
Second, both the mean, $\langle u_i\rangle$ and the fluctuation, $\tilde u_i$, are decomposed into a temporal mean and fluctuation, and as a result,
$$
\frac{\partial}{\partial x_j}\overline{\langle u_iu_j\rangle} =
\frac{\partial}{\partial x_j}[\langle\overline u_i\rangle\langle\overline u_j\rangle + 
\overline{\langle u''_i\rangle\langle u''_j\rangle} + 
\langle\overline{\tilde u_i\tilde u_j}\rangle] 
$$
Third, the decomposition of the spatial fluctuation, $\tilde u_i$ into a temporal mean, $\overline u''_i$, and a temporal fluctuation, $u''_i$, ({\em i.e.} $\tilde u_i = \overline u''_i + u''_i$) results into the final form
$$
\frac{\partial}{\partial x_j}\overline{\langle u_iu_j\rangle} =
\frac{\partial}{\partial x_j}[\langle\overline u_i\rangle\langle\overline u_j\rangle + 
\overline{\langle u''_i\rangle\langle u''_j\rangle} + 
\langle\overline u''_i\overline u''_j\rangle] +
\overline{\langle u''_i u''_j\rangle}]. 
$$
Finally, using the average mass conservation, 
$$
\frac{\partial}{\partial x_j}\overline{\langle u_iu_j\rangle} =
\langle\overline u_j\rangle\frac{\partial\langle\overline u_i\rangle }{\partial x_j}
+
\frac{\partial}{\partial x_j}[
\overline{\langle u''_i\rangle\langle u''_j\rangle}  +
\overline{\langle u''_i u''_j\rangle} + 
\langle\overline u''_i\overline u''_j\rangle]. 
$$
Note that the above decomposition tells us what physical features of the problem has been modelled and the assumptions made to neglect some of the other physics. Similar approach has been used to derive Eq~(\ref{eq:mnc}). 
\end{appendices}

\section*{acknowledgements}
  The author acknowledges financial support from the National Science and Research Council~(NSERC), Canada. Useful discussions with Prof. Jan Martin Nordbotten (jan.nordbotten@math.uib.no) is also greatly acknowledged. Many thanks to two anonymous reviewers for very useful comments and suggestions. The computational work was done partially with a Dell T7400 Workstation funded by the Industrial Research and Innovation Fund~(IRIF), Govt of Newfoundland and Labrador, and partially on the ACEnet (www.ace-net.ca) high performance computing cluster.
%
%

\bibliographystyle{spbasic}
\bibliography{alamjhms2014}

\end{document}